\newif\ifconfver
\confverfalse      

\ifconfver
     \documentclass[10pt,twocolumn,twoside]{IEEEtran}
\else
    \documentclass[11pt,draftcls,onecolumn]{IEEEtran}
\fi

\usepackage{calc,amsfonts,amssymb,amsmath,bm,url,color,theorem,graphicx,cite}
\usepackage{psfrag,subfigure,float}
\usepackage{algorithm2e,multirow,subfigure,amsmath,epsfig,amsfonts,amssymb,psfig,graphics,psfrag,theorem,calc,subfigure,url,bm,cite}
\usepackage{stfloats,hyperref}  

    \makeatletter
    \def\multilimits@{\bgroup
  \Let@
  \restore@math@cr
  \default@tag
 \baselineskip\fontdimen10 \scriptfont\tw@
 \advance\baselineskip\fontdimen12 \scriptfont\tw@
 \lineskip\thr@@\fontdimen8 \scriptfont\thr@@
 \lineskiplimit\lineskip
 \vbox\bgroup\ialign\bgroup\hfil$\m@th\scriptstyle{##}$\hfil\crcr}
    \def\Sb{_\multilimits@}
    \def\endSb{\crcr\egroup\egroup\egroup}
\makeatother

{\begin{list}{}%
    {\setlength{\rightmargin}{0pc}%
    \setlength{\leftmargin}{1pc} }} %
{\end{list}}

\newlength{\twidth}
\ifconfver
\else
\fi

\definecolor{orange}{RGB}{255,107,0}

\newtheorem{Fact}{Fact}

\newtheorem{Prop}{Proposition}
\newtheorem{Theorem}{Theorem}

\theorembodyfont{\rmfamily}

\newtheorem{Remark}{Remark}

\newenvironment{Ventry}[1]%
{\begin{list}{}{
    \settowidth{\labelwidth}{\mbox{\textnormal{#1}}}%
    \setlength{\leftmargin}{\labelwidth+\labelsep}}}%
{\end{list}}

\newcommand{\Rbb}{\mathbb{R}}

\newcommand{\conv}{{\sf conv}}
\newcommand{\aff}{{\sf aff}}
\newcommand{\intr}{\mathsf{int}}
\newcommand{\MVES}{\mathsf{MVES}}

\newcommand{\vca}{\mathbf{a}}
\newcommand{\vcb}{\mathbf{b}}

\newcommand{\vce}{\mathbf{e}}
\newcommand{\vch}{\mathbf{h}}
\newcommand{\vcp}{\mathbf{p}}
\newcommand{\vcs}{\mathbf{s}}
\newcommand{\vct}{\mathbf{t}}
\newcommand{\vcu}{\mathbf{u}}

\newcommand{\vcx}{\mathbf{x}}

\newcommand{\setA}{\mathcal{A}}
\newcommand{\setB}{\mathcal{B}}

\newcommand{\setH}{\mathcal{H}}
\newcommand{\setI}{\mathcal{I}}
\newcommand{\setO}{\mathcal{O}}
\newcommand{\setP}{\mathcal{P}}

\newcommand{\setT}{\mathcal{T}}
\newcommand{\setR}{\mathcal{R}}
\newcommand{\setU}{\mathcal{U}}

\newcommand{\setX}{\mathcal{X}}

\begin{document}

\bibliographystyle{IEEEtran}

\title{A Fast Hyperplane-Based Minimum-Volume Enclosing Simplex Algorithm for Blind Hyperspectral Unmixing}


\ifconfver \else {\linespread{1.1} \rm \fi

\ifconfver
\author{Chia-Hsiang Lin$^*$, Chong-Yung Chi, Yu-Hsiang Wang, and Tsung-Han Chan
\thanks{This work is supported partly by the National Science Council, R.O.C., under
Grant NSC 102-2221-E-007-035-MY2, and partly by Ministry of Science and Technology, R.O.C., under Grant MOST 104-2221-E-007-069-MY3.
Part of this work was presented at 2015 IEEE ICASSP \cite{Lin2015icasspHyperCSI}.
}
\thanks{Chia-Hsiang Lin$^*$ (the corresponding author), Chong-Yung Chi and Yu-Hsiang Wang are with the Institute of Communications Engineering and the Department of Electrical Engineering, National Tsing Hua University, Hsinchu, Taiwan 30013, R.O.C. E-mail: chiahsiang.steven.lin@gmail.com; cychi@ee.nthu.edu.tw; s101064505@m101.nthu.edu.tw.}
\thanks{Tsung-Han Chan is with MediaTek Inc., National Science Park, Hsinchu, Taiwan 30013, R.O.C. E-mail: chantsunghan@gmail.com.}
}

\else
\author{Chia-Hsiang Lin$^*$, Chong-Yung Chi, Yu-Hsiang Wang, and Tsung-Han Chan
\thanks{This work is supported partly by the National Science Council, R.O.C., under
Grant NSC 102-2221-E-007-035-MY2, and partly by Ministry of Science and Technology, R.O.C., under Grant MOST 104-2221-E-007-069-MY3.
Part of this work was presented at 2015 IEEE ICASSP \cite{Lin2015icasspHyperCSI}.
}
\thanks{Chia-Hsiang Lin$^*$ (the corresponding author), Chong-Yung Chi and Yu-Hsiang Wang are with the Institute of Communications Engineering and the Department of Electrical Engineering, National Tsing Hua University, Hsinchu, Taiwan 30013, R.O.C. E-mail: chiahsiang.steven.lin@gmail.com; cychi@ee.nthu.edu.tw; s101064505@m101.nthu.edu.tw.}
\thanks{Tsung-Han Chan is with MediaTek Inc., National Science Park, Hsinchu, Taiwan 30013, R.O.C. E-mail: chantsunghan@gmail.com.}
}
\fi

\maketitle

\ifconfver \else \vspace{-0.5cm}\fi

\begin{abstract}
Hyperspectral unmixing (HU) is a crucial signal processing procedure to identify the underlying materials (or endmembers) and their corresponding proportions (or abundances) from an observed hyperspectral scene.
A well-known blind HU criterion, advocated by Craig in early 1990's, considers the vertices of the minimum-volume enclosing simplex of the data cloud as {good} endmember estimates, and it has been empirically and theoretically found effective even in the scenario of no pure pixels.
However, such kind of algorithms may suffer from heavy simplex volume computations in numerical optimization, etc.
In this work, without involving any simplex volume computations, by exploiting a convex geometry fact that a simplest simplex of $N$ vertices can be defined by $N$ associated hyperplanes, we propose a fast blind HU algorithm, for which each of the $N$ hyperplanes associated with the Craig's simplex of $N$ vertices is constructed from $N-1$ affinely independent data pixels, together with an endmember identifiability analysis for its performance support.
Without resorting to numerical optimization, the devised algorithm searches for the $N(N-1)$ {active} data pixels via simple linear algebraic computations, accounting for its computational efficiency.
Monte Carlo simulations and real data experiments are provided to demonstrate its superior efficacy over some benchmark Craig-criterion-based algorithms in both computational efficiency and estimation accuracy.
\end{abstract}

\ifconfver \else \vspace{-0.0cm}\fi

\ifconfver \else \vspace{-0.5cm}\fi

\ifconfver \else } \fi


\vspace{-0.4cm}

\section{Introduction}\label{sec:introduction}
Hyperspectral remote sensing (HRS) \cite{keshava2002spectral,bioucas13overview, 14SPM}, also known as imaging spectroscopy, is a crucial technology to the identification of material substances (or endmembers) as well as their corresponding fractions (or abundances) present in a scene of interest from observed hyperspectral data, having various applications such as planetary exploration, land mapping and classification, environmental monitoring, and mineral identification and quantification
\cite{Landgrebe2002, Shaw2002, Stein2002}.
The observed pixels in the hyperspectral imaging data cube are often spectral mixtures of multiple substances, the so-called {\it mixed pixel} phenomenon \cite{Jose12}, owing to the limited spatial resolution of the hyperspectral sensor (usually equipped on board the satellite or aircraft) utilized for recording the electromagnetic scattering patterns of the underlying materials in the observed hyperspectal scene
over about several hundreds of narrowly spaced (typically, 5-10 nm) wavelengths that contiguously range from visible to near-infrared bands.
{Occasionally, the mixed pixel phenomenon
can result from the underlying materials intimately mixed \cite{johnson2007snapshot}.}
{\it Hyperspectral unmixing} (HU) \cite{Jose12, Ken14SPM_HU}, an essential procedure of extracting individual spectral signatures of the underlying materials in the captured scene from these measured spectral mixtures, is therefore of paramount importance in the HRS context.

Blind HU, or unsupervised HU, involves two core stages, namely
endmember extraction and abundance estimation, without (or with very limited) prior knowledge about the endmembers' nature or the mixing mechanism. 
Some endmember extraction algorithms (EEAs), such as
alternating projected subgradients (APS) \cite{Zymnis2007},
joint Bayesian approach (JBA) \cite{Nicolas2009}, and
iterated constrained endmembers (ICE) \cite{Berman2004} (also the sparsity promoting ICE (SPICE) \cite{zare2007sparsity}),
can simultaneously determine the associated abundance fractions while extracting the endmember signatures.
Nevertheless, some EEAs perform endmember estimation, followed by abundance estimation using such as the fully constrained least squares (FCLS) \cite{Heinz2001} to complete the entire HU processing.

The pure-pixel assumption has been exploited in devising fast blind HU algorithms to search for the purest pixels over the data set as the endmember estimates, and such searching procedure can always be carried out through simple linear algebraic formulations; see, e.g., pixel purity index (PPI) \cite{Boardman1995} and vertex component analysis (VCA) \cite{Nascimento2005}.
An important blind HU criterion, called Winter's criterion \cite{Winter1999},
also based on the pure-pixel assumption, is to identify the vertices of the maximum-volume simplex inscribed in the observed data cloud as endmember estimates.
HU algorithms in this category include
N-finder (N-FINDR) \cite{Winter1999},
simplex growing algorithm (SGA) \cite{chang2006newnew} (also the real-time implemented SGA \cite{chang2010real}), and
worst case alternating volume maximization (WAVMAX) \cite{chan2011simplex}, to name a few.
However, the pure-pixel assumption could be seriously infringed in practical scenarios especially when the pixels are intimately mixed, for instance, the hyperspectral imaging data for retinal analysis in the ophthalmology \cite{johnson2007snapshot}.
In these scenarios, HU algorithms in this category could completely fail; actually, it is proven that perfect endmember identifiability is impossible
for Winter-criterion-based algorithms if the pure-pixel assumption is violated \cite{chan2011simplex}.

Without relying on the existence of pure pixels, another promising blind HU approach, advocated by Craig in early 1990's \cite{Craig1994}, exploits the simplex structure of hyperspectral data, and believes that the vertices of the minimum-volume data-enclosing simplex can yield {good} endmember estimates, and algorithms developed accordingly include such as
minimum-volume transform (MVT) \cite{Craig1994},
minimum-volume constrained nonnegative matrix factorization (MVC-NMF) \cite{miao2007endmember}, and
minimum-volume-based elimination strategy (MINVEST) \cite{hendrix2012new}.
Moreover, some linearization-based methods have also been reported to practically identify Craig's minimum-volume simplex, e.g.,
the iterative linear approximation in minimum-volume simplex analysis (MVSA) \cite{Li2008} {(also its fast implementation using the interior-point method \cite{CVX2004}, termed as ipMVSA \cite{li2015minimumFastMVSA})}, and
the alternating linear programming in minimum-volume enclosing simplex (MVES) \cite{chan2009convex}.
Empirical evidences do well support that this minimum-volume approach
is resistant to lack of pure pixels, and can recover ground truth endmembers quite accurately even when the observed pixels are heavily mixed.
Very recently, the validity of this empirical belief has been theoretically justified; specifically, we show that, as long as a key measure concerning the pixels' mixing level is above a certain (small) threshold, Craig's simplex can perfectly identify the true endmembers in the noiseless scenario \cite{lin2014identifiability}.
However, 
without the guidance of the pure-pixel assumption, this more sophisticated criterion would generally lead to more computationally expensive HU algorithms.
{To the best of our knowledge, the ipMVSA algorithm \cite{li2015minimumFastMVSA} and the simplex identification via split augmented Lagrangian (SISAL) algorithm \cite{bioucas2009variable} are the two state-of-the-art Craig-criterion-based algorithms in terms of computational efficiency.}
Nevertheless, in view of not only the NP-hardness of the Craig-simplex-identification (CSI) problem \cite{packer2002np} but also heavy simplex volume computations, all the above mentioned HU algorithms are yet to be much more computationally efficient.
Moreover, their performances may not be very reliable owing to the sensitivity to regularization parameter tuning, non-deterministic (i.e., non-reproducible) endmember estimates caused by random initializations, and, most seriously, lack of rigorous identifiability analysis.
%

In this work, we break the deadlock on the trade-off between a simple fast algorithmic scheme and the estimation accuracy in the no pure-pixel case.
We have observed that when the pure-pixel assumption holds true, the effectiveness of a simple fast HU algorithmic scheme
could be attributed to that the desired solutions (i.e., pure pixels) already exist in the data set.
Inspired by this observation, we naturally raise a question: {\em Can Craig's minimum-volume simplex be identified by simply searching for {a specific set of} pixels in the data set regardless of the existence of pure pixels?}
The answer is {affirmative} and will be given in this paper.

Based on the convex geometry fact that a simplest simplex of $N$ vertices can be characterized by the $N$ associated hyperplanes, this paper proposes an efficient and effective unsupervised Craig-criterion-based HU algorithm, together with an endmember identifiability analysis.
Each hyperplane, parameterized by a normal vector and an inner product constant \cite{CVX2004}, can then be estimated from $N-1$ affinely independent pixels in the data set via simple linear algebraic formulations.
The resulting hyperplane-based CSI (HyperCSI) algorithm, based on the above pixel search scheme, can withstand the no pure-pixel scenario, and can yield deterministic, non-negative, and, most importantly, {accurate} endmember estimates.
After endmember estimation, a closed-form expression in terms of the identified hyperplanes' parameters is derived for abundance estimation.
Then some Monte Carlo numerical simulations and real hyperspectral data experiments are presented to demonstrate the superior efficacy of the proposed HyperCSI algorithm over some benchmark Craig-criterion-based HU algorithms in both estimation accuracy and computational efficiency.

The remaining part of this paper is organized as follows.
In Section \ref{sec:prob_setup}, we briefly review some essential convex geometry concepts, followed by 
the signal model and dimension reduction. 
Section \ref{sec:HyperCSI} focuses on the HyperCSI algorithm development, and
in Section \ref{sec:simulation}, some simulation results are presented for its performance comparison with some benchmark Craig-criterion-based HU algorithms.
In Section \ref{sec:RealData}, we further evaluate the effectiveness of the proposed HyperCSI algorithm {with AVIRIS \cite{AVIRISrealdata}} data experiments.
Finally, we draw some conclusions in Section \ref{sec:Conclusions}.

The following notations will be used in the ensuing presentation.
%
%
$\mathbb{R}$ ($\mathbb{R}^N$, $\mathbb{R}^{M\times N}$) is the set of real numbers ($N$-vectors, $M\times N$ matrices).
$\mathbb{R}_{+}$ ($\mathbb{R}^N_{+}$, $\mathbb{R}^{M\times N}_{+}$) is the set of non-negative real numbers ($N$-vectors, $M\times N$ matrices).
$\mathbb{R}_{++}$ ($\mathbb{R}^N_{++}$, $\mathbb{R}^{M\times N}_{++}$) is the set of positive real numbers ($N$-vectors, $M\times N$ matrices).
${\bf X}^{\dag}$ denotes the Moore-Penrose pseudo-inverse of a matrix ${\bf X}$.
${\bf 1}_N$ and ${\bf 0}_N$ are all-one and all-zero $N$-vectors, respectively. 
${\bf e}_i$ denotes the unit vector of proper dimension with the $i$th entry equal to unity.
${\bf I}_N$ is the $N\times N$ identity matrix.
$\succeq$ and $\succ$ stand for the componentwise inequality and strictly componentwise inequality, respectively.
$\|\cdot\|$ denotes the Euclidean norm.
{The distance of a vector $\bf v$ to a set $\cal S$ is denoted by
${\rm dist}({\bf v},{\cal S})\triangleq \inf_{{\bm v}\in{\cal S}} \|{\bf v}-{\bm v}\|$ \cite{CVX2004}.}
{$|{\cal S}|$ denotes the cardinality of the set $\cal S$.}
{The determinant of matrix ${\bf X}$ is represented by ${\rm det}({\bf X})$.}
${\cal I}_{ Z}$ stands for the set of integers $\{1,\dots,{ Z}\}${, for any positive integer $Z$}.


\section{Convex Geometry and Signal Model}
\label{sec:prob_setup}


In this section, a brief review on some essential convex geometry will be given for ease of later use. Then the signal model for representing the hyperspectral imaging data together with dimension reduction preprocessing will be presented.

\vspace{-0.3cm}

\subsection{Convex Geometry Preliminary}
\label{Sec:concept}

\vspace{-0.1cm}

\begin{figure}[t]
\begin{center}
\psfrag{R3}[Bc][Bc]{\Large $\mathbb{R}^3$}
\psfrag{a0}[Bc][Bc]{\Large ${\bf a}_0$}
\psfrag{a1}[Bc][Bc]{\Large ${\bf a}_1$}
\psfrag{a2}[Bc][Bc]{\Large ${\bf a}_2$}
\psfrag{a3}[Bc][Bc]{\Large ${\bf a}_3$}
\psfrag{b}[Bc][Bc]{\Large ${\bf b}$}
\psfrag{af}[Bc][Bc]{\Large $\aff \{ {\bf a}_0, {\bf a}_1 \}$}
\psfrag{afff}[Bc][Bc]{\Large $\aff \{ {\bf a}_1, {\bf a}_2 , {\bf a}_3 \}$ $\equiv \mathcal{H}({\bf b},h)$ }
\psfrag{add}[Bc][Bc]{\Large $\triangleq \{ {\bf x}\in \mathbb{R}^3 ~\big|~ {\bf b}^T {\bf x} = h \}$}
\psfrag{con}[Bc][Bc]{\Large $\conv \{ {\bf a}_0, {\bf a}_3 \}$}
\psfrag{conv}[Bc][Bc]{\Large $\conv \{ {\bf a}_1, {\bf a}_2 , {\bf a}_3 \}$}

\ifconfver
    \resizebox{0.8\linewidth}{!}{\includegraphics{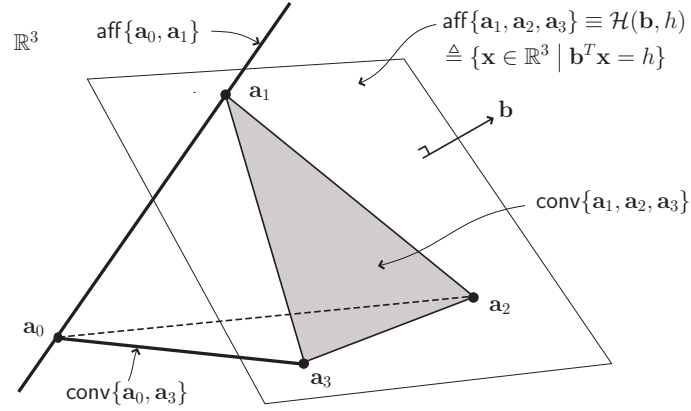}}\vspace{-0.0cm}
\else
    \resizebox{0.5\linewidth}{!}{\includegraphics{concepts.eps}}
\fi
\end{center}
\caption{A graphical illustration in $\mathbb{R}^3$ for some convex geometry concepts, where the line segment connecting ${\bf a}_0$ and ${\bf a}_3$ is the convex hull of $\{ {\bf a}_0, {\bf a}_3 \}$, the straight line passing ${\bf a}_0$ and ${\bf a}_1$ is the affine hull of $\{ {\bf a}_0, {\bf a}_1 \}$, the shaded triangle is the convex hull of $\{ {\bf a}_1,{\bf a}_2,{\bf a}_3 \}$, and the plane passing the three points $\{ {\bf a}_1,{\bf a}_2,{\bf a}_3 \}$ is the affine hull of $\{ {\bf a}_1,{\bf a}_2,{\bf a}_3 \}$. As an affine hull in $\mathbb{R}^3$ is called a hyperplane if its affine dimension is $2$, $\aff\{ {\bf a}_1,{\bf a}_2,{\bf a}_3 \}$ is a hyperplane, while $\aff \{ {\bf a}_0,{\bf a}_1 \}$ is not.} \label{Fig:Basic concept}
\vspace{-0.4cm}
\end{figure}

The {\it convex hull} of a given set of vectors $\{ {\bf a}_1,\ldots, {\bf a}_N \} \subseteq \mathbb{R}^M$ is defined as \cite{CVX2004}
\vspace{-0.2cm}
\begin{equation*}
{\rm conv}\{ {\bf a}_1,\ldots, {\bf a}_N \} \triangleq
\bigg\{  {\bf x} = \sum_{i=1}^N \theta_i {\bf a}_i ~ \bigg| ~
\boldsymbol{\theta} \in \mathbb{R}^N_+, {\bf 1}_N^T\bm{\theta}=1 
\bigg\} ,\label{conv_hull}
\vspace{-0.2cm}
\end{equation*}
where $\boldsymbol{\theta}=[\theta_1,\ldots,\theta_N]^T$ (cf. Figure \ref{Fig:Basic concept}).
A convex hull ${\rm conv}\{ {\bf a}_1,\ldots, {\bf a}_N \}$ is called an $(N-1)$-dimensional {\it simplex} with $N$ vertices $\{{\bf a}_1,\ldots, {\bf a}_N\}$ if $\{{\bf a}_1,\ldots, {\bf a}_N\}$ is {\it affinely independent}, or, equivalently, if $\{ {\bf a}_1 - {\bf a}_N, \ldots, {\bf a}_{N-1} - {\bf a}_N \}$ is linearly independent, and it is called a {\it simplest simplex} in $\mathbb{R}^M$ when $M = N-1$ {\cite{ambikapathi2011chance}}.
For example, a triangle is a $2$-dimensional simplest simplex in $\mathbb{R}^2$, and a tetrahedron is a $3$-dimensional simplest simplex in $\mathbb{R}^3$ (cf. Figure \ref{Fig:Basic concept}).
%

For a given set of vectors $\{ {\bf a}_1,\ldots, {\bf a}_N \} \subseteq \mathbb{R}^M$, its {\it affine hull} is defined as \cite{CVX2004}
\begin{equation*}
{\rm aff}\{ {\bf a}_1,\ldots, {\bf a}_N \} \triangleq
\bigg\{  {\bf x} = \sum_{i=1}^N \theta_i {\bf a}_i ~ \bigg| ~
\boldsymbol{\theta} \in \mathbb{R}^N, {\bf 1}_N^T\bm{\theta}=1 
\bigg\} , \label{eqn: affine hull}
\end{equation*}
where $\boldsymbol{\theta}=[~\theta_1,\ldots,\theta_N~]^T$ (cf. Figure \ref{Fig:Basic concept}). 
This affine hull can be parameterized by a 2-tuple $({\bf C},{\bf d}) \in \mathbb{R}^{M\times P} \times\mathbb{R}^M$ using the following alternative representation \cite{CVX2004}:
\begin{equation*}
{\rm aff}\{ {\bf a}_1,\ldots, {\bf a}_N \} \equiv {\cal A}({\bf C},{\bf d}) \triangleq \big\{ ~ {\bf x}= {\bf C}\boldsymbol{\alpha} + {\bf d} ~ \big| ~ \boldsymbol{\alpha} \in\mathbb{R}^P ~ \big\},\label{eqn:dr_alpha}
\end{equation*}
where $P \triangleq {\rm rank}({\bf C})$ (the rank of $\bf C$) is the {\it affine dimension} of ${\rm aff}\{ {\bf a}_1,\ldots, {\bf a}_N \}$. Moreover, an affine hull ${\rm aff}\{ {\bf a}_1,\ldots, {\bf a}_N \}\subseteq \mathbb{R}^M$ is called a {\it hyperplane} if its affine dimension $P = M-1$ (cf. Figure \ref{Fig:Basic concept}).

\subsection{Signal Model and Dimension Reduction}

Consider a scenario where a hyperspectral sensor measures solar electromagnetic radiation over $M$ spectral bands from $N$ unknown materials (endmembers) in a scene of interest.
{Based on the linear mixing model (LMM) \cite{keshava2002spectral,bioucas13overview,14SPM,Landgrebe2002,Shaw2002,Stein2002,Jose12, Ken14SPM_HU,chan2009convex},
where the measured solar radiations are assumed to reflect from the explored scene through one single bounce,
and the endmembers' spectral signature vectors ${\bf a}_i\in{\mathbb R}^M$ are assumed to be invariant with the pixel index $n$,
each pixel ${\bf x}[n] \in\mathbb{R}^M$ in the observed data set can then be represented as a linear mixture of the $N$ endmembers' spectral signatures}
{\footnote{Note that there is a research line considering non-linear mixtures for modeling the effect of multiple reflections of solar radiation \cite{dobigeon2014nonlinear}.
Moreover, the endmember spectral signatures may be spatially varying, hence leading to the full-additivity in {\sf (A2)} being violated \cite{Ken14SPM_HU}.
However, studying these effects is out of the scope of this paper, and the representative LMM is sufficient for our analysis and algorithm development; interested readers are referred to the magazine papers \cite{dobigeon2014nonlinear} and \cite{zare2014endmembervariability}, respectively, for the non-linear effect and the endmember variability effect.}}
\vspace{-0.1cm}
\begin{equation}\label{eq:sig1}
{\bf x}[n]={\bf A}{\bf s}[n]=\sum^N_{i=1}s_i[n]{\bf a}_i,~~{ \forall n\in\setI_L},
\vspace{-0.1cm}
\end{equation}
where ${\bf A}=[{\bf a}_1\cdots {\bf a}_N]\in \mathbb{R}^{M\times N}$ is the spectral signature matrix, ${\bf s}[n]= [ s_1[n]\cdots s_N[n]]^T\in\mathbb{R}^N$ is the abundance vector, and $L$ is the total number of pixels.
The following standard assumptions pertaining to the model in \eqref{eq:sig1}, which also characterize the simplex structure inherent in the hyperspectral data, are used in our HU algorithm development later
\cite{keshava2002spectral,bioucas13overview,14SPM,Landgrebe2002,Shaw2002,Stein2002,Jose12, Ken14SPM_HU,chan2009convex}:
\vspace{0.06cm}
\begin{Ventry}{}
\item [$\textsf{(A1)}$] (Non-negativity) $s_i[n]\geq 0$, $\forall~i\in\setI_N$ and
$\forall~n\in\setI_L$.

\vspace{0.06cm}

\item [$\textsf{(A2)}$] (Full-additivity) $\sum_{i=1}^N s_i[n] = 1$, $\forall~n\in\setI_L$.

\vspace{0.06cm}

\item [$\textsf{(A3)}$] min$\{L, M\} \geq N$ and ${\bf A}\in \mathbb{R}_{+}^{M\times N}$ is
full
column
rank.
\end{Ventry}
\vspace{0.06cm}
Moreover, like most benchmark HU algorithms (see, e.g., \cite{miao2007endmember,Li2008,chan2009convex,bioucas2009variable}), 
the number of endmembers $N$ is assumed to be known a priori, which can be determined beforehand by applying model-order selection methods, such as
hyperspectral signal subspace identification by minimum error (HySiMe) \cite{Bioucas-Dias2008}, and
Neyman-Pearson detection theory-based virtual dimensionality (VD) \cite{Chang2004}.

We aim to blindly estimate the unknown endmembers (i.e., ${\bf a}_1,\ldots,{\bf a}_N$), as well as their abundances (i.e., ${\bf s}[1],\ldots,{\bf s}[L]$), from the observed spectral mixtures (i.e., ${\bf x}[1],\ldots,{\bf x}[L]$).
Due to the huge dimensionality $M$ of hyperspectral data, directly analyzing the data may not be very computationally efficient. Instead, 
an efficient data preprocessing technique, called affine set fitting (ASF) procedure \cite{Chan2007}, can be applied to compactly represent each measured pixel ${\bf x}[n]\in\mathbb{R}^M$ in a dimension-reduced (DR) space $\mathbb{R}^{N-1}$ as follows:
%
%
\begin{equation}
\tilde{{\bf x}}[n]={\bf C}^{\dag}({\bf x}[n]-{\bf
d}) = \sum_{i=1}^{N}s_i[n] \boldsymbol{\alpha}_i \in\mathbb{R}^{N-1},\label{eq:pprrox}
\vspace{-0.2cm}
\end{equation}
%
%
\noindent where
%
%
\begin{align} 
{\bf d} & = \frac{1}{L} \sum_{n=1}^L {\bf x}[n] \in \mathbb{R}^M ~\text{(mean of data set)} \label{sol:da}\\
{\bf C} & = [~ \boldsymbol{q}_1( {\bf U}{\bf U}^T ),
 \hdots, \boldsymbol{q}_{N-1}(
{\bf U}{\bf U}^T ) ~] \in \mathbb{R}^{M\times (N-1)}\label{sol:Ca}\\
\boldsymbol{\alpha}_i & = {\bf C}^{\dag}({\bf a}_i-{\bf d})\in\mathbb{R}^{N-1}~\text{(DR endmembers)}\label{eq:ai_recover}
\end{align}
in which
${\boldsymbol{q}_i({\bf U}{\bf U}^T)}\in\mathbb{R}^M$ denotes the $i$th principal eigenvector (with unit norm) of the square matrix ${\bf U}{\bf U}^T\in\mathbb{R}^{M\times M}$, and
${\bf U} = [~{\bf x}[1] - {\bf d}, \hdots, {\bf x}[L] - {\bf d} ~] \in \mathbb{R}^{M\times L}$ is the mean-removed data matrix.
Actually, like other dimension reduction algorithms \cite{fodor2002survey}, ASF also performs noise suppression in the meantime.
It has been shown that the above ASF best represents the measured data in an $(N-1)$-dimensional space in the sense of least-squares fitting error \cite{Chan2007}, while such fitting error vanishes in the noiseless scenario \cite{Chan2007}.
Note that the data mean in the DR space is the origin ${\bf 0}_{N-1}$ (by \eqref{eq:pprrox} and \eqref{sol:da}).

Because of $N-1\ll M$ in typical HU applications, the HyperCSI algorithm will be developed in the DR space $\mathbb{R}^{N-1}$ wherein the DR endmembers $\boldsymbol{\alpha}_1,\ldots,\boldsymbol{\alpha}_N$ are estimated.
Then, by \eqref{eq:ai_recover}, the endmember estimates in the original space $\mathbb{R}^M$ can be restored as
%
%
\begin{equation}\label{eq:20140908012422}
\hat{{\bf a}}_i = {\bf C} ~ \hat{\boldsymbol{\alpha}}_i + {\bf d},~~{ \forall~i\in \setI_N},
\end{equation}
where $\hat{\bm{\alpha}}_i$'s are the endmember estimates in the DR space.


\section{Hyperplane-Based Craig-Simplex-Identification Algorithm}
\label{sec:HyperCSI}

First of all, due to \eqref{eq:pprrox} and {\sf (A1)}-{\sf (A2)}, the true endmembers' convex hull ${\sf conv}\{ \bm{\alpha}_1,\ldots, \bm{\alpha}_N \}$ itself is a data-enclosing simplex, i.e.,

\vspace{-0.35cm}

\begin{equation}\label{eq:enclosing_in_DR}
\setX \triangleq \{~\tilde{{\bf x}}[1],\ldots,\tilde{{\bf x}}[L]~\} \subseteq {\sf conv}\{\boldsymbol{\alpha}_1,\ldots, \boldsymbol{\alpha}_N\}.
\end{equation}
According to Craig's criterion, the true endmembers' convex hull is estimated by minimizing the volume of the data-enclosing simplex \cite{Craig1994}, namely, by solving the following volume minimization problem (called the CSI problem interchangeably hereafter):

\vspace{-0.3cm}

\begin{equation}\label{eq:MESA_pprrob}
\begin{split}
 \min\begin{Sb}{\bm{\beta}_i \in\mathbb{R}^{N-1},\forall i }\end{Sb}~~~ &V(\bm{\beta}_1, \ldots
,\bm{\beta}_N)\\
 {\rm s.t.}~~~ &\tilde{{\bf x}}[n]\in{\sf conv}
\{\boldsymbol{\beta}_1,\ldots, \boldsymbol{\beta}_N \},~\forall n,
\end{split}
\end{equation}
where $V(\bm{\beta}_1, \ldots,\bm{\beta}_N)$ denotes the volume of the simplex ${\sf conv}\{ \bm{\beta}_1,\ldots, \bm{\beta}_N \}\subseteq \mathbb{R}^{N-1}$. Under some mild conditions on data purity level \cite{lin2014identifiability}, the optimal solution of problem \eqref{eq:MESA_pprrob} can perfectly yield the true endmembers $\bm{\alpha}_1,\ldots, \bm{\alpha}_N$.
{\footnote{In \cite{lin2014identifiability}, $\widetilde{\gamma} \triangleq \max\{ r'\leq 1 \mid \mathcal{T}_e \cap \mathcal{B}(r') \subseteq {\rm conv}\{ {\bf s}[1], \ldots, {\bf s}[L] \} \}$ is used to measure the data purity level of $\setX$,
where ${\cal T}_e\triangleq {\rm conv}\{{\bf e}_1,\ldots,{\bf e}_N\}\subseteq\mathbb{R}^N$ and
${\cal B}(r')\triangleq\{ {\bf x}\in\mathbb{R}^N\mid \|{\bf x}\|\leq r' \}$; the geometric
interpretations of $\widetilde{\gamma}$ can be found in \cite{lin2014identifiability}.
Simply speaking, one can show that $\widetilde{\gamma}\in[\frac{1}{\sqrt{N}},1]$, and the most heavily mixed scenario (i.e., ${\bf s}[n]= \frac{ 1 }{N}{\bf 1}_N,~\forall n\in\setI_L$) will lead to the lower bound \cite{lin2014identifiability}.
On the contrary, the pure-pixel assumption is equivalent to the condition of $\widetilde{\gamma}=1$ (the upper bound) \cite{lin2014identifiability}, comparing to which a mild condition of only $\widetilde{\gamma}>\frac{1}{\sqrt{N-1}}$ is sufficient to guarantee the perfect endmember identifiability of problem \eqref{eq:MESA_pprrob} \cite{lin2014identifiability}.
}}

Besides in the HU context, the NP-hard CSI problem in \eqref{eq:MESA_pprrob} \cite{packer2002np} has been studied in some earlier works in mathematical geology \cite{Full81} and computational geometry \cite{Zhou2002}.
However, their intractable computational complexity almost disable them from practical applications for larger problem size \cite{Zhou2002}, mainly owing to calculation of the complicated {\it nonconvex} objective function \cite{chan2009convex}

\vspace{-0.4cm}

\[
V(\bm{\beta}_1, \ldots,\bm{\beta}_N) 
=
\frac{1}{(N-1)!}\cdot 
{\left|{\rm det}\left(
\left[
\begin{array}{ccc}
                \bm{\beta}_1 & \cdots & \bm{\beta}_N \\
                1 & \cdots & 1 \\
              \end{array}
            \right]
\right)\right|}
\vspace{-0.05cm}
\]
in \eqref{eq:MESA_pprrob}. 
Instead, the HyperCSI algorithm to be presented can judiciously bypass 
simplex volume calculations, and meanwhile the identified simplex can be shown to be exactly the ``minimum-volume" (data-enclosing) simplex in the asymptotic sense ($L\rightarrow \infty$).

\begin{figure}[h]
\begin{center}
  \psfrag{R2}[Bc][Bc]{\LARGE $\mathbb{R}^2$}
  \psfrag{a1}[Bc][Bc]{\LARGE $\boldsymbol{ \alpha}_1$ }
  \psfrag{a2}[Bc][Bc]{\LARGE $\boldsymbol{ \alpha}_2$ }
  \psfrag{a3}[Bc][Bc]{\LARGE $\boldsymbol{ \alpha}_3$ }
  \psfrag{b1}[Bc][Bc]{\LARGE $\vcb_1$ }
  \psfrag{b2}[Bc][Bc]{\LARGE $\tilde{\vcb}_1$ }
  \psfrag{b3}[Bc][Bc]{\LARGE $\hat{\vcb}_1$ }  
  \psfrag{c11}[Bc][Bc]{\LARGE $\tilde{\boldsymbol{ \alpha}}_1$ }
  \psfrag{c1}[Bc][Bc]{\LARGE $\tilde{\boldsymbol{ \alpha}}_2$ }
  \psfrag{c2}[Bc][Bc]{\LARGE $\tilde{\boldsymbol{ \alpha}}_3$ }  
  \psfrag{H}[Bc][Bc]{\LARGE $\widetilde{\setH}_1$ }
  \psfrag{Hh}[Bc][Bc]{\LARGE $\widehat{\setH}_1$ }  
  \psfrag{G}[Bc][Bc]{\LARGE ${\setH}_1$ } 
  \psfrag{xn}[Bc][Bc]{\LARGE $\tilde{\vcx}[n]$ } 
  \psfrag{O}[Bc][Bc]{\LARGE $ {\bf 0}_2$}
  \psfrag{p1}[Bc][Bc]{\LARGE $\vcp_1^{(1)}$ }
  \psfrag{p2}[Bc][Bc]{\LARGE $\vcp_2^{(1)}$ }
  \psfrag{r11}[Bc][Bc]{\LARGE ${\cal R}_1^{(1)}$  }
  \psfrag{r21}[Bc][Bc]{\LARGE ${\cal R}_2^{(1)}$  }
  \psfrag{q}[Bc][Bc]{\LARGE $ \bf{q}$ }
  
  \ifconfver
      \resizebox{0.9\linewidth}{!}{\includegraphics{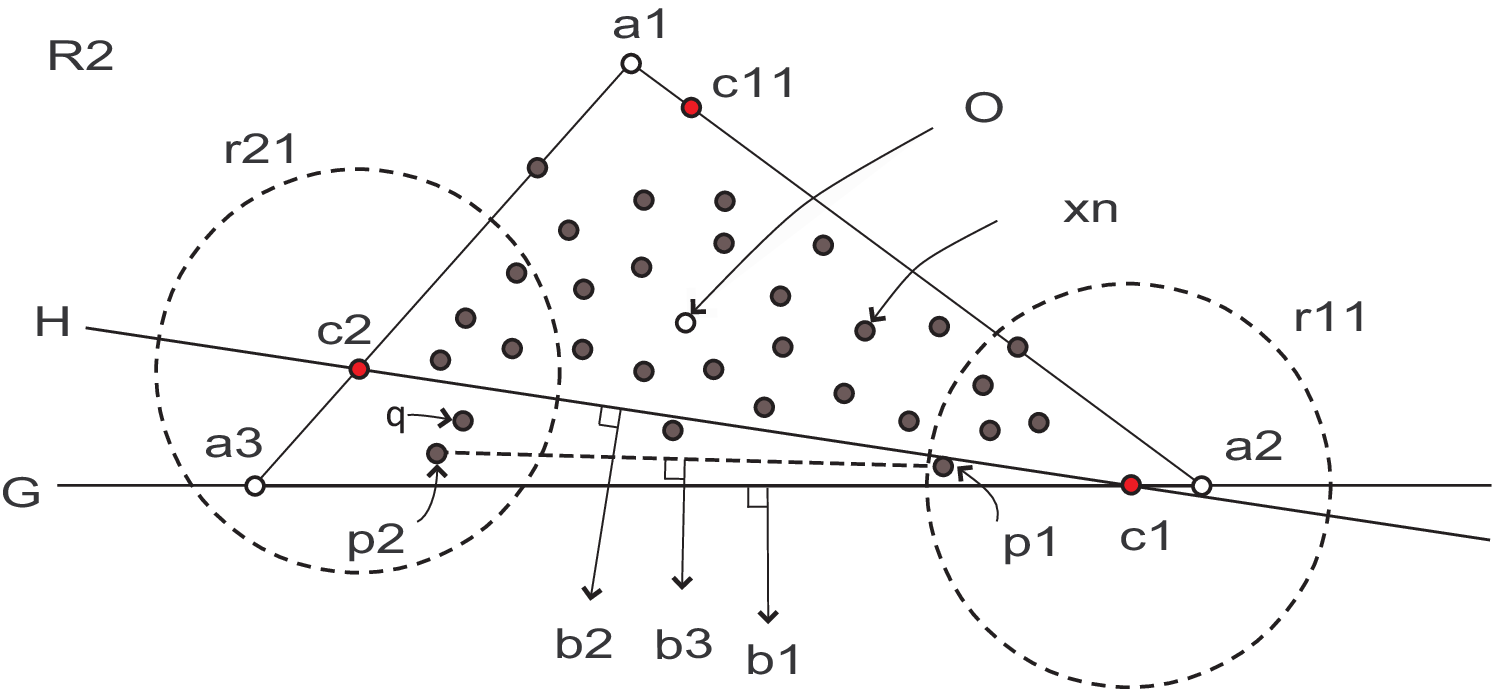}}
  \else
      \resizebox{0.7\linewidth}{!}{\includegraphics{ICASSP2015_fig_new3.eps}}
  \fi

\end{center}
\vspace{-0.5cm}
\caption{An illustration of hyperplanes and DR data in ${\mathbb R}^2$ for the case of $N=3$, where $\tilde{\boldsymbol{ \alpha}}_3$ is a purest pixel in $\setX$ (a purest pixel $\tilde{\boldsymbol{ \alpha}}_i$ can be considered as the pixel closest to ${\boldsymbol{ \alpha}}_i$) but not necessarily very close to  hyperplane $\setH_1={\sf aff}\{ \boldsymbol{ \alpha}_2,\boldsymbol{ \alpha}_3 \}$, leading to nontrivial orientation difference between $\tilde{\vcb}_1$ and $\vcb_1$.
However, the {active} pixels $\vcp_1^{(1)}$ and $\vcp_2^{(1)}$ identified by \eqref{eq:multiple_vector} will be very close to $\setH_1$ (especially, for large $L$), and hence the orientations of $\hat{\bf b}_1$  and $\vcb_1$ will be almost the same.
{On the other hand, one can see that the pixels identified by \eqref{eq:multiple_vector_simple} are $\{\vcp_2^{(1)},{\bf q}\}$ (that are very close to each other) whose corresponding normal vector estimate is obviously far away from the true $\vcb_1$.}
}\label{Fig:ICASSP2015_11111111}
\vspace{-0.4cm}
\end{figure}

{First of all, let us succinctly present the actual idea on which the HyperCSI algorithm is based.
%
As the Craig's minimum-volume simplex can be uniquely determined by $N$ tightly enclosed $(N-2)$-dimensional hyperplanes, where each hyperplane can be reconstructed from $N-1$ affinely independent points on itself, we hence endeavor to search for $N-1$ affinely independent pixels (referred to as {\it active} pixels in $\setX$) that are as close to the associated hyperplane as possible.
We begin with $N$ purest pixels that define $N$ disjoint proper regions, each centered at a different purest pixel.
Then for each hyperplane of the minimum-volume simplex, the desired $N-1$ active pixels, that are as close to the hyperplane as possible, are respectively sifted from $N-1$ subsets of $\setX$, each enclosed in one different proper region (cf. Figure \ref{Fig:ICASSP2015_11111111}).
Then the obtained $N-1$ pixels are used to construct one estimated hyperplane.
Finally, the desired minimum-volume simplex can be determined from the obtained $N$ hyperplane estimates. 
}

\vspace{-0.4cm}

\subsection{Hyperplane Representation for Craig's Simplex}

{The idea of solving the CSI problem in \eqref{eq:MESA_pprrob}, without involving any simplex volume computations, is based on the hyperplane representation of a simplest simplex as stated in the following proposition:}

\vspace{-0.4cm}

\begin{Prop}\label{proposition:equ_rep}
If $\{\boldsymbol{\alpha}_1,\ldots, \boldsymbol{\alpha}_N\} \subseteq\Rbb^{N-1}$ is affinely independent, i.e.,
$\setT = {\sf conv}\{\boldsymbol{\alpha}_1,\ldots, \boldsymbol{\alpha}_N\} \subseteq\Rbb^{N-1}$ is a simplest simplex,
then $\setT$ can be reconstructed from the associated $N$ hyperplanes $\{\setH_1,\ldots,\setH_N\}$, that tightly enclose $\setT$, where $\setH_i \triangleq {\sf aff} (~ \{ \boldsymbol{\alpha}_1,\ldots, \boldsymbol{\alpha}_N \} \setminus \{ \boldsymbol{\alpha}_i \} ~)$.
\end{Prop}
\emph{Proof:}
It suffices to show that $\{\boldsymbol{\alpha}_1,\ldots, \boldsymbol{\alpha}_N\}$ can be determined by $\{\setH_1,\ldots,\setH_N\}$. It is known that hyperplane $\setH_i$ can be parameterized by a normal vector ${\bf b}_i \in \mathbb{R}^{N-1}$ and an inner product constant $h_i\in\mathbb{R}$ as follows:
\begin{equation}\vspace{-0.1cm}
\setH_i(\vcb_i,h_i) = \big\{ ~ {\bf x}\in\mathbb{R}^{N-1} ~ \big| ~  {\bf b}_i^T{\bf x} = h_i ~ \big\}.\label{eqn:prop1_1}
\end{equation}
As $\boldsymbol{\alpha}_i \in  {\sf aff} (~ \{ \boldsymbol{\alpha}_1,\ldots, \boldsymbol{\alpha}_N \} \setminus \{ \boldsymbol{\alpha}_j \} ~) = \setH_j$ for all $j\neq i$, we have from \eqref{eqn:prop1_1} that ${\bf b}_j^T \boldsymbol{\alpha}_i = h_j$ for all $j\neq i$, i.e.,
\begin{equation}
{\bf B}_{-i} \boldsymbol{\alpha}_i = \vch_{-i},\label{eqn:prop1_2}
\end{equation}
where ${\bf B}_{-i}\in \Rbb^{(N-1)\times(N-1)}$, $\vch_{-i}\in\Rbb^{(N-1)}$ are defined as
\begin{align}\vspace{-0.15cm}
{\bf B}_{-i} &\triangleq [\vcb_1,\dots,\vcb_{i-1},\vcb_{i+1},\dots, \vcb_N]^T \label{B-i},\\
\vch_{-i} &\triangleq [h_1,\dots, h_{i-1}, h_{i+1}, \dots, h_N]^T \label{h-i}.
\end{align}
As $\setT$ is a simplest simplex in $\mathbb{R}^{N-1}$, ${\bf B}_{-i}$ must be of full rank and hence invertible \cite{CVX2004}.
Hence, we have from \eqref{eqn:prop1_2} that
\begin{equation}
\boldsymbol{\alpha}_i={\bf B}_{-i}^{-1}~\vch_{-i},~\forall~i\in\setI_N,\label{eqn:prop1_3}
\end{equation}
implying that $\boldsymbol{\alpha}_i$ can be reconstructed.
The proof is therefore completed.\hfill$\blacksquare$

As it can be inferred from {\sf (A3)} that the set of DR endmembers $\{\boldsymbol{\alpha}_1,\ldots, \boldsymbol{\alpha}_N\}$ is affinely independent,
one can apply Proposition \ref{proposition:equ_rep} to decouple the CSI problem \eqref{eq:MESA_pprrob} into $N$ subproblems of hyperplane estimation, namely, estimation of $N$ parameter vectors $({\bf b}_i,h_i)$ in \eqref{eqn:prop1_1}. 
Then \eqref{eqn:prop1_3} can be utilized to obtain the desired endmember estimates.
Next, let us present how to estimate the normal vector $\vcb_i$ and the inner product constant $h_i$ from the data set $\setX$, respectively.

\subsection{Normal Vector Estimation}


The normal vector $\vcb_i$ of hyperplane $\setH_i$ can be obtained by projecting the vector $\boldsymbol{\alpha}_j - \boldsymbol{\alpha}_i$ (for any $j\neq i$) onto the subspace that is orthogonal to the hyperplane $\setH_i$ \cite{friedberglinear1997}, i.e.,
\vspace{-0.1cm}
\begin{align}
\vcb_i &\equiv {\bm v}_i( \boldsymbol{\alpha}_1,\ldots,\boldsymbol{\alpha}_N)  \label{eq:formula_of_bi}\\
& \triangleq \left( {\bf I}_{N-1} - {\bf P} ({\bf P}^T {\bf P})^{-1} {\bf P}^T  \right) \cdot (\boldsymbol{\alpha}_j - \boldsymbol{\alpha}_i),~\text{for any}~j\neq i,\nonumber
\vspace{-0.1cm}
\end{align}
where ${\bf P} \triangleq {\bf Q} - \boldsymbol{\alpha}_j\cdot {\bf 1}_{N-2}^T \in \mathbb{R}^{(N-1)\times(N-2)}$, and ${\bf Q} \in \mathbb{R}^{(N-1)\times(N-2)}$ is the matrix $[\boldsymbol{\alpha}_1 \cdots \boldsymbol{\alpha}_N] \in \mathbb{R}^{(N-1)\times N}$ with its $i$th and $j$th columns removed.
Besides \eqref{eq:formula_of_bi} for obtaining the normal vector ${\bf b}_i$ of $\setH_i$, we also need another closed-form expression of ${\bf b}_i$ in terms of $N-1$ distinct points as given in the following proposition.

\vspace{-0.3cm}

\begin{Prop}\label{proposition:bi_formula_with_zero}
Given any affinely independent set $ \{\vcp_{1}^{(i)},\ldots,\vcp_{N-1}^{(i)}\}\subseteq \setH_i$,
${\bf b}_i$ can be alternatively obtained by (except for a positive scale factor)
\begin{equation}
{\vcb}_i = {\bm v}_i( \vcp_{1}^{(i)},\ldots,\vcp_{i-1}^{(i)}, {\bf 0}_{N-1}, \vcp_{i}^{(i)},\ldots, \vcp_{N-1}^{(i)} ),\label{eq:formula_of_bi_hat}
\end{equation}
where ${\bm v}_i(\cdot)$ is defined in \eqref{eq:formula_of_bi}. 
\end{Prop}

\vspace{-0.2cm}

\noindent{The proof of Proposition \ref{proposition:bi_formula_with_zero} can be shown from the fact that ${\bf 0}_{N-1}$ is the data mean in the DR space ${\mathbb R}^{N-1}$ (by \eqref{eq:pprrox} and \eqref{sol:da}), and is omitted here
due to space limitation.}

Based on Proposition \ref{proposition:bi_formula_with_zero}, we estimate the normal vector $\vcb_i$ by finding $N-1$ affinely independent data points
%
%
\begin{equation}
\setP_i \triangleq
\{\vcp_{1}^{(i)},\ldots,\vcp_{N-1}^{(i)}\}  \subseteq\setX \nonumber
\end{equation}
that are as close to $\setH_i$ as possible.
To this end, an observation from \eqref{eq:enclosing_in_DR} is needed and given in the following fact:

\vspace{-0.25cm}

\begin{Fact}\label{fact:0219}
{Observing that (i) $\vcb_i^T\vcp\leq h_i,$ $\forall\vcp\in\setX$ (i.e., all the points $\vcp\in\setX$ lie on the same side of $\setH_i$; cf. \eqref{eq:enclosing_in_DR}), and that (ii) ${\rm dist}({\bf p},{\setH_i}) = |h_i-\vcb_i^T\vcp| / \|\vcb_i\|$, the point} $\vcp\in \setX$ closest to $\setH_i$ is exactly the one with maximum of $\vcb_i^T\vcp$, provided that $\vcb_i$ points outward from the true endmembers' simplex (cf. Figure \ref{Fig:ICASSP2015_11111111} and \eqref{eq:formula_of_bi}).
\end{Fact}
\vspace{-0.2cm}

Suppose that we are given $N$ ``purest" pixels $\tilde{\boldsymbol{\alpha}}_i \in \setX$, which basically maximize the simplex volume inscribed in $\setX$, and they can be obtained using the reliable and reproducible successive projection algorithm (SPA) \cite{Ken14SPM_HU}, \cite[Algorithm 4]{SPAarora2012practical}.
So $\tilde{\boldsymbol{\alpha}}_i$ can be viewed as the pixel in $\setX$ ``closest" to $\boldsymbol{\alpha}_i$ (cf. Figure \ref{Fig:ICASSP2015_11111111}).
Let $\tilde{\vcb}_i$ be the outward-pointing normal vector of hyperplane $\widetilde{\setH}_i \triangleq {\sf aff} (~ \{ \tilde{\boldsymbol{\alpha}}_1,\ldots, \tilde{\boldsymbol{\alpha}}_N \} \setminus \{ \tilde{\boldsymbol{\alpha}}_i \} ~)$, i.e.,
\begin{align}
\tilde{\vcb}_i &\triangleq {\bm v}_i (\tilde{\boldsymbol{\alpha}}_1,\ldots,\tilde{\boldsymbol{\alpha}}_N).~~\text{(cf. \eqref{eq:formula_of_bi})} \label{eq:bi_tilde_def}
\end{align}
Considering Fact \ref{fact:0219} and the requirement that {the set $\setP_i$ must contain $N-1$ distinct elements} (otherwise, $\setP_i$ is not affinely independent), we identify the desired affinely independent set $\setP_i$ by:
\begin{equation}
\vcp^{(i)}_k~\in~\arg\max~ \{{\tilde{\vcb}_i^T}\vcp \mid \vcp\in \setX \cap \setR^{(i)}_k \},~~\forall~k\in\setI_{N-1},\label{eq:multiple_vector}
\end{equation}
where $\setR_1^{(i)},\ldots,\setR_{N-1}^{(i)}$ are $N-1$ disjoint sets defined as
\vspace{-0.15cm}
\begin{align}
\setR_k^{(i)} \equiv \setR_k^{(i)}(\tilde{\boldsymbol{\alpha}}_1,\ldots,\tilde{\boldsymbol{\alpha}}_N) &\triangleq \begin{cases}
\setB(\tilde{\boldsymbol{\alpha}}_k,r),~k<i,\\
\setB(\tilde{\boldsymbol{\alpha}}_{k+1},r),~k\geq i,
\end{cases} \label{eq:region_def}
\vspace{-0.15cm}
\end{align}
in which $\setB(\tilde{\boldsymbol{\alpha}}_k,r) \triangleq \{ \vcx\in\mathbb{R}^{N} \mid \|\vcx - \tilde{\boldsymbol{\alpha}}_k\| < r \}$ is the open Euclidean norm ball with center $\tilde{\boldsymbol{\alpha}}_k\in\mathbb{R}^{N}$ and radius $r \triangleq ({ 1 }/{2})\cdot \min\{ \| \tilde{\boldsymbol{\alpha}}_i - \tilde{\boldsymbol{\alpha}}_j \| \mid 1\leq i<j\leq N \}>0$.
Note that the choice of the radius $r$ is to guarantee that $\setR^{(i)}_1,\ldots,\setR^{(i)}_{N-1}$ are $N-1$ non-overlapping regions, thereby guaranteeing that $\setP_i$ contains $N-1$ distinct points.
{Moreover, each hyperball $\setR^{(i)}_k$ must contain at least one pixel (as it contains either $\tilde{\boldsymbol{\alpha}}_k$ or $\tilde{\boldsymbol{\alpha}}_{k+1}$; cf. \eqref{eq:region_def}), i.e., $\setX \cap \setR^{(i)}_k\neq \emptyset$, and hence problem \eqref{eq:multiple_vector} must be a feasible problem (i.e., a problem with non-empty feasible set \cite{CVX2004}).
}

If the $N-1$ points extracted by \eqref{eq:multiple_vector} are affinely independent, then the estimated normal vector associated with $\setH_i$ can be determined as (cf. Proposition \ref{proposition:bi_formula_with_zero})
\vspace{-0.1cm}
\begin{equation}\label{eq:formula_of_bi_hat_hat}
\hat{{\vcb}}_i = {\bm v}_i( \vcp_{1}^{(i)},\ldots,\vcp_{i-1}^{(i)}, {\bf 0}_{N-1}, \vcp_{i}^{(i)},\ldots, \vcp_{N-1}^{(i)} ).
\vspace{-0.1cm}
\end{equation}
Fortunately, the obtained $\setP_i$ by \eqref{eq:multiple_vector} can be proved (in Theorem \ref{prop:AI} below) to be always affinely independent with one more assumption:
{\footnote{The rationale of adopting Dirichlet distribution in {\sf (A4)} is not only that it is a well known distribution that captures both the non-negativity and full-additivity of ${\bf s}[n]$ \cite{frigyik2010introduction}, but because it has been used to characterize the distribution of ${\bf s}[n]$ in the HU context \cite{nascimento2012hyperspectral,nascimento2007hyperspectral}.
However, the statistical assumption {\sf (A4)} is only for analysis purpose
without being involved in our geometry-oriented algorithm development.
So even if abundance vectors are neither i.i.d. nor Dirichlet distributed, the HyperCSI algorithm can still work well; cf. Subsection \ref{subsec:all_algo_2new_maps}.
Furthermore, we would like to emphasize that, in our analysis (Theorems \ref{prop:AI} and \ref{thm:identifiability}), we actually only use the following two properties of Dirichlet distribution: (i) its domain is $\mathsf{dom}~f = \{\vcs\in\mathbb{R}^N_{++} \mid {\bf 1}^T_N\vcs = 1\}$, and (ii) it is a continuous multivariate distribution with strictly positive density on its domain \cite{johnson2002continuous}; cf. Appendixes \ref{proofsec:AI} and \ref{proofsec:THM_identifiability}.
Hence, any distribution with these two properties can be used as an alternative in {\sf (A4)}.
}}
\begin{Ventry}{}
\item [$\textsf{(A4)}$] The abundance vectors $\{\vcs[n]\} \subseteq \mathbb{R}^N$ {(defined below \eqref{eq:sig1})} are independent and identically distributed (i.i.d.) following Dirichlet distribution with parameter $\boldsymbol{\gamma} = [\gamma_1,\ldots,\gamma_N] \in \mathbb{R}^N_{++}$ whose probability density function (p.d.f.) is given by \cite{frigyik2010introduction}:  
\begin{equation}\label{eq:dirich_pdf}
f(\vcs) = \begin{cases}
\frac{\Gamma(\gamma_0)}{\prod_{i=1}^N \Gamma(\gamma_i) } \cdot \prod_{i=1}^N s_i^{\gamma_i -1}, & \vcs\in \mathsf{dom}~f,\\
0, & \textrm{otherwise},
\end{cases}
\vspace{-0.15cm}
\end{equation}
where $\vcs = [s_1,\ldots,s_N] \in \mathbb{R}^N$, $\gamma_0 = \sum_{i=1}^N \gamma_i$, $\mathsf{dom}~f = \{\vcs\in\mathbb{R}^N_{++} \mid {\bf 1}^T_N\vcs = 1\}$, and $\Gamma(\gamma) = \int_0^\infty x^{\gamma-1} e^{-x}~dx$ denotes the gamma function.
\end{Ventry}

\vspace{-0.35cm}

\begin{Theorem}\label{prop:AI}
Assume {\sf (A1)}-{\sf (A4)} hold true. Let $\vcp_k^{(i)}\in\setP_i$ be a solution to  \eqref{eq:multiple_vector} with $\setR_k^{(i)}$ defined in \eqref{eq:region_def}, for all $i\in\setI_N$ and $k\in\setI_{N-1}$.
Then, the set $\setP_i$ is affinely independent with probability $1$ (w.p.1).
\end{Theorem}

\vspace{-0.2cm}

\noindent The proof of Theorem \ref{prop:AI} is given in Appendix \ref{proofsec:AI}.

Note that the orientation difference between $\tilde{\vcb}_i$ and the true ${\vcb}_i$ may not be small (cf. Figure \ref{Fig:ICASSP2015_11111111}).
Hence, $\tilde{\vcb}_i$ itself may not be a good estimate for ${\vcb}_i$ either.
On the contrary, it can be shown that the orientation difference between $\hat{\vcb}_i$ and ${\vcb}_i$ tends to be small for large $L$, and actually such difference vanishes as $L$ goes to infinity (cf. Theorem \ref{thm:identifiability} as well as Remark \ref{rmk:thm2} in {Subsection \ref{sec:ComplexityAnalysis}}).
{On the other hand, if the pixels with maximum inner products in $\setP_i$ are jointly sifted from the whole data cloud $\setX$, i.e.,
\begin{equation}
\setP_i~\in~\arg\max~ \{{\tilde{\vcb}_i^T}(\vcp_1+\cdots +\vcp_{N-1}) \mid \setP\subseteq \setX
\}, \label{eq:multiple_vector_simple}
\end{equation}
where $\setP\triangleq \{ \vcp_1,\ldots,\vcp_{N-1} \}$,
rather than respectively from different regions $\setX\cap \setR_k^{(i)}$, $\forall k\in\setI_{N-1}$, as given in \eqref{eq:multiple_vector}, the identified pixels in $\setP_i$ may stay quite close, easily leading to large deviation in normal vector estimation as illustrated in Figure \ref{Fig:ICASSP2015_11111111} where $\setP_i=\{ \vcp_2^{(1)},{\bf q} \}$ are the identified pixels using \eqref{eq:multiple_vector_simple}. This is also a rationale of finding $\setP_i$ using \eqref{eq:multiple_vector} for better normal vector estimation.
}


\subsection{Inner Product Constant Estimation}\label{InnerProductConstantEstimation}

For Craig's simplex (the minimum-volume data-enclosing simplex), all the data in $\setX$ should lie on the same side of $\setH_i$ (otherwise, it is not data-enclosing), and $\setH_i$ should be as tightly close to the data cloud $\setX$ as possible (otherwise, it is not minimum-volume); the only possibility is when the hyperplane $\setH_i$ must be externally tangent to the data cloud.
In other {words}, $\setH_i$ will incorporate the pixel that has maximum inner product with $\hat{\vcb}_i$, and hence it can be determined as $\setH_i(\hat{\vcb}_i,\hat{h}_i)$, where $\hat{h}_i$ is obtained by solving
%
%
\begin{equation}
\hat{h}_i~=~\max~\{~ {\hat{\vcb}_i^T}\vcp \mid \vcp\in \setX ~\}.\label{eq:compute_inner_product_constant_hat}
\end{equation}

However, it has been reported that when the observed data pixels are noise-corrupted, the random noise may expand the data cloud, thereby inflating the volume of the Craig's data-enclosing simplex \cite{chan2011simplex,ambikapathi2011chance}.
As a result, the estimated hyperplanes are pushed away from the origin (i.e., the data mean in the DR space) due to noise effect, and hence the estimated inner product constant in \eqref{eq:compute_inner_product_constant_hat} would be larger than that of the ground truth.
To mitigate this effect, the estimated hyperplanes need to be properly shifted closer to the origin, so instead, $\setH_i(\hat{\vcb}_i,\hat{h}_i/c)$, $\forall~i\in\setI_N$, are the desired hyperplane estimates for some $c\geq 1$. Therefore, the corresponding DR endmember estimates are obtained by (cf. \eqref{eqn:prop1_3})
%
%
\begin{equation}
\hat{\boldsymbol{\alpha}}_i = \widehat{{\bf B}}_{-i}^{-1} \cdot \frac{\hat{\vch}_{-i}}{c} ,~~~\forall~i\in\setI_N,\label{eqn:estimated_endmembers}
\end{equation}
where $\widehat{{\bf B}}_{-i}$ and $\hat{\vch}_{-i}$ are given by \eqref{B-i} and \eqref{h-i} with ${\bf b}_j$ and $h_j$ replaced by $\hat{\bf b}_j$ and $\hat{h}_j$, $\forall~ j\neq i$, respectively. Moreover, it is necessary to choose $c$ such that the associated endmember estimates in the original space are non-negative (cf. {\sf (A3)}), i.e.,
%
%
\begin{equation}\vspace{-0.0cm}
\hat{{\bf a}}_i = {\bf C} ~ \hat{\boldsymbol{\alpha}}_i + {\bf d}\succeq {\bf 0}_M,~\forall~i\in\setI_N. ~~\text{(cf. \eqref{eq:20140908012422})}\label{eq:201409080124}
\end{equation}
By \eqref{eqn:estimated_endmembers} and \eqref{eq:201409080124}, the hyperplanes should be shifted closer to the origin
with $c =c'$ at least, where
%
%
\begin{equation}
c' ~\triangleq~\min_{c''\geq 1} \{ c'' \mid {\bf C} ~( \widehat{{\bf B}}_{-i}^{-1} \cdot \hat{\vch}_{-i} ) + c''\cdot {\bf d} \succeq {\bf 0}_M,~\forall~i \}\label{eq:cprime_def} 
\end{equation}
which can be further shown to have a closed-form solution:
\begin{equation}
c' = \max\big\{ 1 ,  \max \{ -{v}_{ij}/d_j   \mid   i\in\setI_N,~ j\in\setI_M \}  \big\}, \label{eq:cprime_closed_form}
\end{equation}
where ${v}_{ij}$ is the $j$th component of ${\bf C} ~( \widehat{{\bf B}}_{-i}^{-1} \cdot \hat{\vch}_{-i} ) \in \mathbb{R}^M$ and $d_j$ is the $j$th component of ${\bf d}$. 

Note that $c'$ is just the minimum value for $c$ to yield non-negative endmember estimates.
Thus, we can generally set $c=c'/\eta \geq c'$ for some $\eta\in(0,1]$.
Moreover, the value of $\eta=0.9$ is empirically found to be a good choice for the scenarios where signal-to-noise ratio (SNR) is greater than $20$ dB; typically, the value of SNR in hyperspectral data is much higher than $20$ dB \cite{AVIRISrealdata}. Let us emphasize that the larger the value of $\eta$ (or the smaller the value of $c$), the farther the estimated hyperplanes from the origin ${\bf 0}_{N-1}$, or the closer the estimated endmembers' simplex ${\sf conv}\{ \hat{\vca}_1,\ldots,\hat{\vca}_N \}$ to the boundary of the nonnegative orthant $\mathbb{R}^M_{+}$.    
On the other hand, we empirically observed that typical endmembers in the U.S. geological survey (USGS) library \cite{USGS2007} are close to the boundary of $\mathbb{R}^M_{+}$. That is to say, a reasonable choice of $\eta \in (0,1]$ should be relatively large (i.e., relatively close to $1$), accounting for the reason 
why the preset value of $\eta = 0.9$ can always yield good performance.
The resulting endmember estimation processing of the HyperCSI algorithm
is summarized in Steps 1 to 6 in Table \ref{table_HyperCSI}.


\ifconfver
\begin{table}[h]
\begin{center}
\caption{Pseudo-code for HyperCSI Algorithm}\label{table_HyperCSI}
\parbox{0.48\textwidth}
{
\rule{\linewidth}{1pt}
\begin{Ventry}{ {\bf Step 1.} }
\item[{\bf Given}]   Hyperspectral data $\{{\bf x}[1],\ldots,{\bf x}[L]\}$, number of endmembers $N$, and $\eta=0.9$.

\item[{\bf Step 1.}] Calculate $({\bf C},{\bf d})$ using \eqref{sol:da}-\eqref{sol:Ca}, and obtain the DR data $\setX = \{ \tilde{{\bf x}}[1],\ldots,\tilde{{\bf x}}[L] \}$ using \eqref{eq:pprrox}.

\item[{\bf Step 2.}] Obtain $\{\tilde{\boldsymbol{\alpha}}_1, \ldots,\tilde{\boldsymbol{\alpha}}_N\}$ using SPA \cite{SPAarora2012practical}.

\item[{\bf Step 3.}] Obtain $\tilde\vcb_i$ using \eqref{eq:bi_tilde_def}, $\forall~i$, and $\setX\cap\setR_k^{(i)}$ using \eqref{eq:region_def}, $\forall~i,k$.

\item[{\bf Step 4.}] Obtain $( \setP_i,\hat{\vcb}_i,\hat{h}_i  )$ by \eqref{eq:multiple_vector}, \eqref{eq:formula_of_bi_hat_hat}, and \eqref{eq:compute_inner_product_constant_hat}, $\forall~i$.

\item[{\bf Step 5.}] Obtain $c'$ by \eqref{eq:cprime_closed_form}, and set $c=c'/\eta$.

\item[{\bf Step 6.}] Calculate $\hat{\boldsymbol{\alpha}}_i$ by \eqref{eqn:estimated_endmembers} and $\hat{{\bf a}}_i = {\bf C} ~ \hat{\boldsymbol{\alpha}}_i + {\bf d}$ by \eqref{eq:201409080124}, $\forall~i$.

\item[{\bf Step 7.}] Calculate $\hat{\bf s}[n] = [\hat{s}_1[n] \cdots \hat{s}_N[n]]^T$ by \eqref{eq:abundance_est_closed_form}, $\forall~n$.

\item[{\bf Output}] The endmember estimates $\{ \hat{{\bf a}}_1,\ldots,\hat{{\bf a}}_N \}$ and abundance estimates $\{ \hat{{\bf s}}[1],\ldots,\hat{{\bf s}}[L] \}$.

\end{Ventry}
\rule{\linewidth}{1pt}

}
\end{center}
\vspace{-0.3cm}
\end{table}
\else
\begin{table}[h]
\begin{center}
\caption{Pseudo-code for HyperCSI Algorithm}\label{table_HyperCSI}
\parbox{0.8\textwidth}
{
\rule{\linewidth}{1pt}
\begin{Ventry}{ {\bf Step 1.} }
\item[{\bf Given}]   Hyperspectral data $\{{\bf x}[1],\ldots,{\bf x}[L]\}$, number of endmembers $N$, and $\eta=0.9$.

\item[{\bf Step 1.}] Calculate $({\bf C},{\bf d})$ using \eqref{sol:da}-\eqref{sol:Ca}, and obtain the DR data $\setX = \{ \tilde{{\bf x}}[1],\ldots,\tilde{{\bf x}}[L] \}$ using \eqref{eq:pprrox}.

\item[{\bf Step 2.}] Obtain $\{\tilde{\boldsymbol{\alpha}}_1, \ldots,\tilde{\boldsymbol{\alpha}}_N\}$ using SPA \cite{SPAarora2012practical}.

\item[{\bf Step 3.}] Obtain $\tilde\vcb_i$ using \eqref{eq:bi_tilde_def}, $\forall~i$, and $\setR_k^{(i)}$ using \eqref{eq:region_def}, $\forall~i,k$.

\item[{\bf Step 4.}] Obtain $( \setP_i,\hat{\vcb}_i,\hat{h}_i  )$ by \eqref{eq:multiple_vector}, \eqref{eq:formula_of_bi_hat_hat}, and \eqref{eq:compute_inner_product_constant_hat}, $\forall~i$.

\item[{\bf Step 5.}] Obtain $c'$ by \eqref{eq:cprime_closed_form}, and set $c=c'/\eta$.

\item[{\bf Step 6.}] Calculate $\hat{\boldsymbol{\alpha}}_i$ by \eqref{eqn:estimated_endmembers} and $\hat{{\bf a}}_i = {\bf C} ~ \hat{\boldsymbol{\alpha}}_i + {\bf d}$ by \eqref{eq:201409080124}, $\forall~i$.

\item[{\bf Step 7.}] Calculate $\hat{\bf s}[n] = [\hat{s}_1[n] \cdots \hat{s}_N[n]]^T$ by \eqref{eq:abundance_est_closed_form}, $\forall~n$.

\item[{\bf Output}] The endmember estimates $\{ \hat{{\bf a}}_1,\ldots,\hat{{\bf a}}_N \}$ and abundance estimates $\{ \hat{{\bf s}}[1],\ldots,\hat{{\bf s}}[L] \}$.

\end{Ventry}
\rule{\linewidth}{1pt}

}
\end{center}
\end{table}
\fi

\subsection{Abundance Estimation}
\label{sec:AbundanceEstimations}

\vspace{0.1cm}

{Though the abundance estimation is often done by solving
FCLS problems \cite{Heinz2001}, which can be equivalently formulated in the DR space as (cf. \cite[Lemma 3.1]{SPUHRDBPS2011})
\vspace{0.1cm}
\begin{equation}\label{eq:FCLS_pprrob_DR}\vspace{-0.15cm}
\begin{split}
 \min\begin{Sb}{{\bf s}' \in\mathbb{R}^{N}}\end{Sb}~~~ &\|\tilde{\bf x}[n]-[\hat{\boldsymbol{\alpha}}_1\cdots\hat{\boldsymbol{\alpha}}_N]{\bf s}'\|\\
 {\rm s.t.}~~~ &{\bf s}'\succeq{\bf 0}_N,~{\bf 1}^T_N{\bf s}'=1,
\end{split}
\vspace{0.22cm}
\end{equation}
it has been reported that some geometric quantities, acquired during the endmember extraction stage, can be used to significantly accelerate the abundance estimation procedure \cite{BarycentricGU2012}.
With similar computational efficiency improvements taken into account, we aim at expressing the abundance $s_i[n]$ in terms of readily available quantities (e.g., normal vectors and inner product constants) obtained when estimating the endmembers, in this subsection.
The results are summarized in the following {proposition}:}
%
%
\begin{Prop}\label{thm:abundance_closed_form}
Assume $\textsf{(A1)}$-$\textsf{(A3)}$ hold true. Then ${\bf s}[n] = [s_1[n] \cdots s_N[n]]^T $ has the following closed-form expression:
%
%
\begin{eqnarray}\label{eq:abundance_closed_form}
s_i[n] = \frac{h_i - {\bf b}_i^T \tilde{{\bf x}}[n] }{h_i - {\bf b}_i^T \boldsymbol{\alpha}_i} ,~~\forall~ i \in \setI_N,~\forall~n \in \setI_L.
%
\vspace{0.1cm}
\end{eqnarray}
\end{Prop}

\noindent {Proposition \ref{thm:abundance_closed_form} can be derived from some simple geometrical observations (cf. items (i) and (ii) in Fact 1) and the following well known formula in the {\it Algebraic Topology} context
\vspace{0.1cm}
\begin{equation}\label{eq:barycentricSN}
s_i[n]=
\frac{{\rm dist}(\tilde{\bf x}[n],\setH_i)}{{\rm dist}(\boldsymbol{\alpha}_i,\setH_i)},
%
\vspace{0.1cm}
\end{equation}
and its proof is omitted here due to space limitation; note that the formula \eqref{eq:barycentricSN} has been recently derived again using different approach in the HU context \cite[Equation (12)]{BarycentricGU2012}.}

Based on \eqref{eq:abundance_closed_form}, the abundance vector ${\bf s}[n] $ can be estimated as
%
%
\begin{eqnarray}\label{eq:abundance_est_closed_form}
\hat{s}_i[n] = \left( \frac{\hat{h}_i - \hat{\bf b}_i^T \tilde{{\bf x}}[n] }{\hat{h}_i - \hat{\bf b}_i^T \hat{\boldsymbol{\alpha}}_i} \right)^{+} ,~~\forall~ i \in \setI_N,~\forall~n \in \setI_L,
\end{eqnarray}
where $(y)^{+}\triangleq \max\{y,0\}$ is to enforce the non-negativity of abundance fractions ${\bf s}[n]$ (cf. Step 7 in Table \ref{table_HyperCSI}).
{One can show that when $\tilde{\bf x}[n]\in {\rm conv}\{\hat{\boldsymbol{\alpha}}_1,\ldots,\hat{\boldsymbol{\alpha}}_N\}$, the abundance estimates obtained using \eqref{eq:abundance_est_closed_form} is exactly the solution to the FCLS problem in \eqref{eq:FCLS_pprrob_DR}, while using \eqref{eq:abundance_est_closed_form} has much lower computational cost than solving FCLS problems.
Nevertheless, one should be aware of a potential limitation of using \eqref{eq:abundance_est_closed_form}.
Specifically, if $\tilde{\bf x}[n]$ is too far away from the endmembers' simplex ${\rm conv}\{\hat{\boldsymbol{\alpha}}_1,\ldots,\hat{\boldsymbol{\alpha}}_N\}$ (i.e., $\hat{\vcb}_i^T \tilde{\bf x}[n]$ is much larger than $\hat{h}_i$ for some $i$), the zeroing operation in \eqref{eq:abundance_est_closed_form} could cause nontrivial deviation in abundance estimation.
This can happen if $\tilde{\bf x}[n]$ is an outlier or the SNR is very low.
However, as the SNR is reasonably high (like in AVIRIS data \cite{chan2011simplex,AVIRISrealdata}), most pixels in the hyperspectral data are expected to lie inside or very close to the endmembers' simplex ${\rm conv}\{\hat{\boldsymbol{\alpha}}_1,\ldots,\hat{\boldsymbol{\alpha}}_N\}$ (cf. {\sf (A1)}-{\sf (A2)})---especially when the endmembers are extracted based on Craig's criterion.
Hence, with the endmembers estimated by the Craig-criterion-based HyperCSI algorithm, simply using \eqref{eq:abundance_est_closed_form} to enforce the abundance non-negativity is not only computationally efficient, but also still capable of yielding good abundance estimation as will be demonstrated 
in the simulation results (Table \ref{tab:pur_snr_end} in Subsection \ref{subsec:all_algo} and Table \ref{tab:pur_snr_end_2newmaps} in Subsection \ref{subsec:all_algo_2new_maps}) later.
}

Unlike most of the existing abundance estimation algorithms, where   
all the $N$ abundance maps must be jointly estimated (e.g., FCLS \cite{Heinz2001}), the proposed HyperCSI algorithm offers an option of 
solely obtaining the abundance map of a specific material of interest (say the $i$th material)
\vspace{-0.15cm}
\begin{equation}\label{eq:def_abudance_map}
{\bm s}_i \triangleq [ s_i[1] \cdots s_i[L] ]^T \in \mathbb{R}_{+}^L,
\vspace{-0.15cm}
\end{equation}
to save computational cost, or obtaining all the abundance maps ${\bm s}_1,\ldots,{\bm s}_N$ by parallel processing (cf. \eqref{eq:abundance_est_closed_form}). 
Moreover, when calculating ${\bm s}_i$ using \eqref{eq:abundance_est_closed_form}, the denominator $\hat{h}_i - \hat{\bf b}_i^T \hat{\boldsymbol{\alpha}}_i$ is a constant for all pixel indices $n\in\setI_L$ and hence only needs to be calculated once regardless of $L$ (which is usually large).

\subsection{Identifiability and Complexity of HyperCSI}
\label{sec:ComplexityAnalysis}


{In this subsection, let us present the identifiability and complexity analyses of the proposed HyperCSI algorithm.}
Particularly, the asymptotic identifiability of the HyperCSI algorithm can be guaranteed as stated in the following theorem with the proof given in Appendix \ref{proofsec:THM_identifiability}:
\vspace{-0.15cm}
\begin{Theorem}\label{thm:identifiability}
Under $\textsf{(A1)}$-$\textsf{(A4)}$, the noiseless assumption and $L\rightarrow \infty$, the simplex identified by HyperCSI algorithm with $c=1$ (in Step 5 in Table \ref{table_HyperCSI}) is exactly the Craig's minimum-volume simplex (i.e., solution of \eqref{eq:MESA_pprrob}) and the true endmembers' simplex ${\sf conv}\{ \boldsymbol{\alpha}_1,\ldots,\boldsymbol{\alpha}_N\}$ w.p.1.
\end{Theorem}
\vspace{-0.15cm}
\noindent Two noteworthy remarks about the philosophies and intuitions behind the proof of this theorem are given as follows:
\vspace{-0.15cm}
\begin{Remark}\label{rmk:thm2}
With the abundance distribution stated in {\sf (A4)}, the $N-1$ pixels in $\setP_i$ can be shown to be arbitrarily close to $\setH_i$ as the pixel number $L\rightarrow \infty$, and they are affinely independent w.p.1 (cf. Theorem \ref{prop:AI}). Therefore, $\hat{\vcb}_i$ can be uniquely obtained by \eqref{eq:formula_of_bi_hat_hat}, and its orientation approaches to that of ${\vcb}_i$ w.p.1.
\vspace{-0.15cm}
\end{Remark}
\begin{Remark}\label{rmk:thm2_hi}
Remark \ref{rmk:thm2} together with \eqref{eq:enclosing_in_DR} implies that $\hat{h}_i$ is upper bounded by $h_i$ w.p.1 (assuming without loss of generality that $\|\hat{\vcb}_i\| = \|\vcb_i\|$), and this upper bound can be shown to be achievable w.p.1 as $L\rightarrow \infty$.
Thus, as $c=1$, we have that $\hat{h}_i/c = h_i$ w.p.1.
\vspace{-0.15cm}
\end{Remark}
It can be further inferred, from the above two remarks, that $\hat{\boldsymbol{\alpha}}_i$ is exactly the true ${\boldsymbol{\alpha}}_i$ w.p.1 (cf. \eqref{eqn:estimated_endmembers}) as $L\rightarrow \infty$ in the absence of noise.
Although the identifiability analysis in Theorem \ref{thm:identifiability} is conducted for the noiseless case and $L\rightarrow \infty$, 
we empirically found that the HyperCSI algorithm can yield {good} endmember estimates for a moderate $L$ and finite SNR, to be demonstrated by simulation results and real data experiments later.

{Next, we analyze the computational complexity of the HyperCSI algorithm.
The computation time of HyperCSI is primarily dominated by the computations of the feasible sets $\setX \cap {\cal R}_k^{(i)}$ (in Step 3), the active pixels in ${\cal P}_i$ (in Step 4), and the abundances $\hat{\bf s}[n]$ (in Step 7), and they are respectively analyzed in the following:
\begin{Ventry}{}
\item [\emph{Step 3:}] Computing the $N(N-1)$ feasible sets $\setX \cap {\cal R}_k^{(i)}$, $\forall i\in \setI_{N}$, $\forall k\in \setI_{N-1}$, is equivalent to computing the $N$ sets $\setX \cap {\cal B}(\tilde{\boldsymbol{\alpha}}_i,r)$, $\forall i\in \setI_{N}$; cf. \eqref{eq:region_def}.
Since ${\cal B}(\tilde{\boldsymbol{\alpha}}_i,r)$ is an open Euclidean norm ball, the computation of each set $\setX \cap {\cal B}(\tilde{\boldsymbol{\alpha}}_i,r)$ can be done by examining $|\setX|=L$ inequalities $\| \tilde{\bf x}[n] - \tilde{\boldsymbol{\alpha}}_i \| <r$, $\forall n\in \setI_{L}$.
However, examining each inequality requires (i) calculating one Euclidean 2-norm (in $\mathbb{R}^{N-1}$), which costs $\setO(N)$,
and (ii) checking whether this 2-norm is smaller than $r$, which costs $\setO(1)$.
Hence, Step 3 costs $\setO(N(N+1)L)$.

\item [\emph{Step 4:}] To determine $\setP_i$, we have to identify the pixel $\vcp_k^{(i)}$ from the set $\setX\cap {\cal R}_k^{(i)}$ ($\forall k\in\setI_{N-1}$), whose complexity amounts to computing $|\setX\cap {\cal R}_k^{(i)} |$ inner products in $\mathbb{R}^{N-1}$ (each costs $\setO(N)$), and performing the point-wise maximum operation among the values of these inner products (cf. \eqref{eq:multiple_vector}), and hence the complexity of identifying $\vcp_k^{(i)}$ is easily verified as $\setO(N\cdot |\setX\cap {\cal R}_k^{(i)} | + |\setX\cap {\cal R}_k^{(i)} | -1)$.
Moreover, gathering $\setP_i= \{\vcp_1^{(i)},\ldots,\vcp_{N-1}^{(i)}\}$ requires the complexity $\sum_{k=1}^{N-1} \setO(N\cdot |\setX\cap {\cal R}_k^{(i)} | + |\setX\cap {\cal R}_k^{(i)} | -1)= \setO( (N+1) \cdot \sum_{k=1}^{N-1}|\setX\cap {\cal R}_k^{(i)} |) \leq \setO((N+1) \cdot |\setX|) =\setO((N+1) L)$; the inequality is due to that ${\cal R}_k^{(i)}$s are disjoint.
Repeating the above for $\setP_i,~\forall i\in\setI_N$, Step 4 costs $\setO(N(N+1)L)$.

\item [\emph{Step 7:}] Estimation of the abundances requires to compute the fraction in \eqref{eq:abundance_est_closed_form} $NL$ times. Each fraction involves $2$ inner products (in $\mathbb{R}^{N-1}$), $2$ scalar subtractions, and $1$ scalar division, and thus costs $\setO(N)$. So, this step costs $\setO(N^2L)$. 
\end{Ventry}
\noindent Therefore, the overall computational complexity of HyperCSI is $\setO( 2N(N+1)L+N^2L )=\setO(N^2 L)$.

Surprisingly, the complexity order $\setO(N^2 L)$ of the proposed HyperCSI algorithm is the same as (rather than much higher than) that of some pure-pixel-based EEAs; see, e.g., \cite{SPAarora2012practical,ambikapathi2011two,Nascimento2005,chan2011simplex}.
Moreover, to the best of our knowledge, the MVES algorithm \cite{chan2009convex} that approximates the CSI problem in \eqref{eq:MESA_pprrob} as alternating linear programming (LP) problems, and solves the LPs using primal-dual interior-point method \cite{CVX2004}, is the existing Craig-criterion-based algorithm with lowest complexity order $\setO(\tau N^2 L^{1.5})$, where $\tau$ is the
number of iterations \cite{chan2009convex}.
Hence, the introduced hyperplane identification approach (without simplex volume computations) indeed yields a smaller complexity than most existing Craig-criterion-based algorithms.}

Let us conclude this section with a summary of some remarkable features of the proposed HyperCSI algorithm (given in Table \ref{table_HyperCSI}) as follows:
\begin{Ventry}{}
\item[(a)] Without involving any simplex volume computations, the Craig's minimum-volume simplex is reconstructed from $N$ hyperplane estimates, i.e., the $N$ estimates $(\hat{\vcb}_i,\hat{h}_i)$, which can be obtained in parallel (cf. Step 4 in Table \ref{table_HyperCSI}) by searching $N(N-1)$ most informative pixels from
$\setX$.
The reconstructed simplex in the DR space $\mathbb{R}^{N-1}$ is actually the intersection of $N$ halfspaces $\{  {\bf x}\in\mathbb{R}^{N-1} \mid  \hat{\bf b}_i^T{\bf x} \leq  \hat{h}_i\},~ i=1,\ldots,N$.

{\item[(b)] By noting that $s_i[n]=0$ if, and only if, ${\bf x}[n]\in\setH_i$, the potential requirement of $N(N-1)$ pixels lying on, or close to,
the associated hyperplanes is considered not difficult to be met in practice because hyperspectral images are often with highly sparse abundances.
This will be discussed in more detail in experiments with AVIRIS data in Section \ref{sec:RealData}.}

\item[(c)] All the processing steps (including SPA in Step 2 of Table \ref{table_HyperCSI}; cf. Algorithm 4 in \cite{SPAarora2012practical}) can be carried out either by simple linear algebraic formulations or by closed-form expressions, and so its high computational efficiency can be anticipated.
\end{Ventry}

\section{Computer Simulations}
\label{sec:simulation}


This section demonstrates the efficacy of the proposed HyperCSI algorithm by Monte Carlo simulations. In the simulation, 
endmember signatures
with $M=224$ spectral bands randomly selected from the USGS library \cite{USGS2007} are used to generate $L$ noise-free synthetic hyperspectral data ${\bf x}[n]$ according to linear mixing model in \eqref{eq:sig1}, where the abundance vectors are i.i.d. and generated following the Dirichlet distribution with $\boldsymbol{\gamma} = {\bf 1}_N/N$ (cf. \eqref{eq:dirich_pdf}) as it can automatically enforce {\sf (A1)} and {\sf (A2)} \cite{chan2009convex,ambikapathi2011chance}. Then we add i.i.d. zero-mean Gaussian noise with variance $\sigma^2$ to the noise-free synthetic data ${\bf x}[n]$ for different values of SNR defined as
SNR$\triangleq ( \sum_{n=1}^L \|{\bf x}[n]\|^2 )/( \sigma^2ML )$,
and those negative entries in the generated noisy data vectors are artificially set to zero, so as to maintain the non-negativity nature of hyperspectral imaging data. 

The root-mean-square (rms) spectral angle error between the true endmembers $\{{{\bf a}}_1,\ldots,{{\bf a}}_N\}$ and their estimates $\{\hat{{\bf a}}_1,\ldots,\hat{{\bf a}}_N\}$ defined as \cite{Nascimento2005,chan2009convex}
\vspace{-0.16cm}
\begin{equation}\label{SSE_a}
\phi_{en}=\min_{\boldsymbol{\pi}\in\Pi_N}\sqrt{\frac{1}{N}\sum_{i=1}^N\left[\arccos\left(\frac{{\bf
a}_i^T\hat{{\bf a}}_{\pi_i}}{\|{\bf a}_i\| \cdot \|\hat{{\bf
a}}_{\pi_i}\|}\right)\right]^2},
\vspace{-0.16cm}
\end{equation}
is used as the performance measure of endmember estimation, where $\Pi_N=\{ \boldsymbol{\pi} = (\pi_1,\ldots,\pi_N) \in \mathbb{R}^N~|~ \pi_i \in \{ 1, \ldots, N \}, ~ \pi_i\neq\pi_j~{\rm for}~i\neq j\}$ is the set of all permutations of $\{1,\ldots ,N\}$.
Similarly, the performance measure of abundance estimation is the rms angle error defined as \cite{chan2009convex}
\vspace{-0.16cm}
\begin{equation}\label{SSE_a_abundace}
\phi_{ab}=\min_{\boldsymbol{\pi}\in\Pi_N}\sqrt{\frac{1}{N}\sum_{i=1}^N\left[\arccos\left(\frac{{\bm s}_i^T\hat{{\bm s}}_{\pi_i}}{\|{\bm s}_i\| \cdot \|\hat{{\bm s}}_{\pi_i}\|}\right)\right]^2},
\vspace{-0.16cm}
\end{equation}
where ${\bm s}_i $ and $\hat {\bm s}_i$ are the true abundance map of $i$th endmember (cf. \eqref{eq:def_abudance_map}) and its estimate, respectively.
All the HU algorithms under test are implemented in Mathworks Matlab R2013a  running on a desktop computer equipped with Core-i7-4790K CPU with 4.00 GHz speed and 16 GB random access memory, and all the performance results in terms of $\phi_{en}$, $\phi_{ab}$, and computational time $T$ are averaged over 100 independent realizations. 

Next, we show some simulation results for 
the endmember identifiability for moderately finite data length (cf. Theorem \ref{thm:identifiability}), 
the choice of the parameter $\eta$, 
and the performance evaluation of the proposed HyperCSI algorithm, in the following subsections, respectively.

\vspace{-0.15cm}

\subsection{Endmember Identifiability of HyperCSI for Finite Data}\label{subsec:thm1}


In Theorem \ref{thm:identifiability},
the perfect endmember identifiability of the proposed HyperCSI algorithm (with $c=1$ in Step 5 in Table \ref{table_HyperCSI}) under the noise-free scenario is proved in the asymptotic sense (i.e., the data length $L\rightarrow\infty$). In this subsection, we would like to show some simulation results to illustrate the asymptotic identifiability of the HyperCSI algorithm and its 
good endmember estimation accuracy even with a moderately finite number of pixels $L$.

\begin{figure}[!]
\psfrag{phi}[Bc][Bc]{\LARGE $\phi_{en}$ (degrees)}
\psfrag{L}[Bc][Bc]{\LARGE number of pixels $L$}
\ifconfver 
\begin{center}
    \resizebox{0.8\linewidth}{!}{\includegraphics{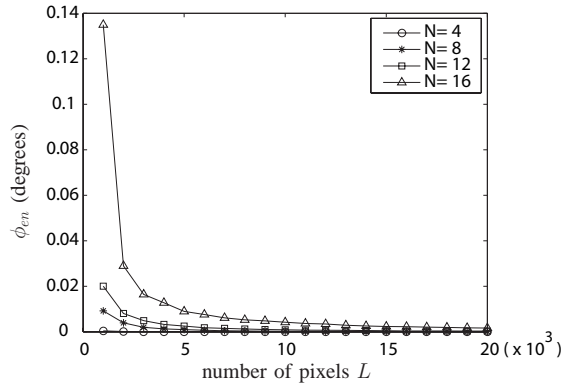}}\vspace{-0.3cm}
\end{center}
\else
\begin{center}
    \resizebox{0.44\linewidth}{!}{\includegraphics{Theorem1_N_16.eps}}
\end{center}
    \vspace{-0.6cm}
\fi 
\caption{The endmember identifiability of the HyperCSI algorithm with finite data length $L$.}
\label{Fig:Theorem1}
\vspace{-0.5cm}
\end{figure}

Figure \ref{Fig:Theorem1} shows some simulation results of $\phi_{en}$ versus $L$ for $N\in\{4,8,12,16\}$. From this figure, one can observe that for a given $N$, $\phi_{en}$ decreases as $L$ increases, and 
the HyperCSI algorithm indeed achieves perfect identifiability (i.e., $\phi_{en}=0$, cf. \eqref{SSE_a}) as $L\rightarrow\infty$.
On the other hand, the HyperCSI algorithm needs to identify $N(N-1)$ essential pixels ${\bf p}_k^{(i)}$ for the construction of the Craig's simplex, which indicates that the HyperCSI algorithm would need more pixels to achieve good performance for larger $N$. 
Intriguingly, the results shown in Figure \ref{Fig:Theorem1} are consistent with the above inferences, where a larger $N$ corresponds to a slightly slower convergence rate of $\phi_{en}$.
However, these results also allude to a high possibility that the HyperCSI algorithm can yield accurate endmember estimates with a typical data length $L$ (i.e., several ten thousands) for high SNR in HRS applications.

\vspace{-0.15cm}

\subsection{Choice of the Parameter $\eta$}\label{subsec:eta}

\begin{figure*}[!]
    \psfrag{eta}[Bc][Bc]{\Large $\eta$}
    \psfrag{phi}[Bc][Bc]{\Large $\phi_{en}$ (degrees)}
    \begin{center}
              \resizebox{1.0\textwidth}{!}{\includegraphics{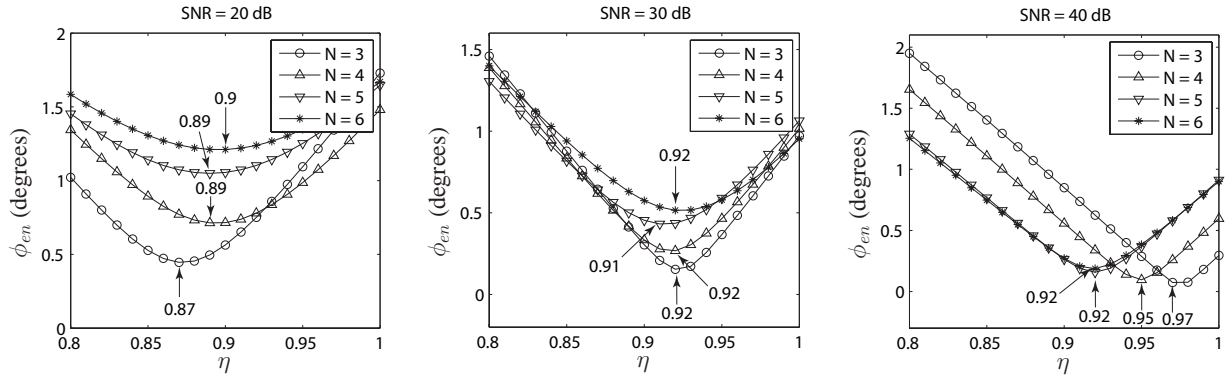}}
          \end{center}
          \vspace{-0.5cm}
\caption{The average r.m.s. spectral angle error $\phi_{en}$ versus different values of $\eta$.
}\label{fig:eta_choise}
\end{figure*}

The simulation results for $\phi_{en}$ versus $\eta$ obtained by the proposed HyperCSI algorithm, for $L=10000$, ${\rm SNR}\in\{20,30,40\}$ (dB), and $N\in\{3,4,5,6\}$ are shown in Figure \ref{fig:eta_choise}. 
From this figure, one can observe that for a fixed $N$, the best choice of $\eta$ (i.e., the one that yields the smallest $\phi_{en}$) decreases as SNR decreases. 
The reason for this is that the larger the noise power, the more the data cloud is expanded, and hence the more the desired hyperplanes should be shifted towards the data center (implying a {larger} $c$ or  a smaller $\eta$).
Moreover, one can also observe from Figure \ref{fig:eta_choise} that for each scenario of $(N,{\rm SNR})$, the best choice of $\eta$ basically belongs to the interval $[0.8,1]$, a relatively large value in the interval $(0,1]$, as discussed in Subsection \ref{InnerProductConstantEstimation}.
It is also interesting to note that for a given SNR, the best choice of $\eta$ tends to approach the value of 0.9 as $N$ increases. For instance, for ${\rm SNR}=20$ dB, the best choices of $\eta$ for $N\in\{3,4,5,6\}$ are $\{0.87,0.89,0.89,0.9\}$, respectively.

The above observations also suggest that $\eta=0.9$, the only parameter in the proposed HyperCSI algorithm, is a good choice. 
Next, we will evaluate the performance of the proposed HyperCSI algorithm with the parameter $\eta$ preset to $0.9$ for all the simulated scenarios and real data tests, though it may not be the best choice for some scenarios.

\vspace{-0.15cm}

\subsection{Performance Evaluation of HyperCSI Algorithm}\label{subsec:all_algo}


We evaluate the performance of the proposed HyperCSI algorithm,
along with a performance comparison with {five} state-of-the-art Craig-criterion-based HU algorithms, including
MVC-NMF \cite{miao2007endmember}, 
MVSA \cite{Li2008},
MVES \cite{chan2009convex}, 
SISAL \cite{bioucas2009variable},
{and ipMVSA \cite{li2015minimumFastMVSA}}.
As the operations of MVC-NMF, MVSA, SISAL, {and ipMVSA} are data-dependent, their respective regularization parameters have been well selected in the simulation, so as to yield their best performances.
In particular, the regularization parameter involved in SISAL is the regression weight for robustness against noise, and hence has also been tuned w.r.t. different SNRs.
The implementation details and parameter settings for all the algorithms under test are listed in Table \ref{parameters_table}.
\begin{table}[ht]
\footnotesize \caption{Simulation settings for the algorithms under test.} \label{parameters_table} \vspace{0.15cm} \centering
{\begin{tabular}{|c|l|}\hline
{\bf{Algorithms}}& {~~~~~\bf{Implementation details and parameter settings}}
\\\hline
               
\multirow{3}*{MVC-NMF}
&{Dimension reduction: Singular value decomposition;}\\%
               &{Regularization parameter: $10^{-3}$;~{Max iteration: $500$;}}\\ 
               & Initialization: VCA-FCLS; Convergence tolerance: $10^{-6}$.\\\hline%

\multirow{2}*{MVSA}
&{Dimension reduction: Principal component analysis;}\\%
               &{Regularization parameter: $10^{-6}$; Initialization: VCA.}\\  \hline%
               
\multirow{2}*{MVES}
               &{Dimension reduction: ASF; Convergence tolerance: $10^{-8}$;}\\%
                              &{Initialization: Solving feasibility problem.}\\ \hline%

\multirow{3}*{SISAL}
&{Dimension reduction: Principal component analysis;}\\%
&{Regularization parameter: $0.015, 0.02, 0.025, 0.03, 0.035$}\\%
               &{w.r.t SNR= $20, 25, 30, 35, 40$ (dB); Initialization: VCA.}\\  \hline%
 
 \multirow{2}*{ipMVSA}
 &{Dimension reduction: Principal component analysis;}\\%
                &{Regularization parameter: $10^{-6}$; Initialization: VCA.}\\  \hline%
               
\multirow{1}*{HyperCSI}
&{Dimension reduction: ASF; $\eta=0.9$.}\\%
\hline%

\end{tabular}}
\end{table}

The {purity index} $\rho_n$ for each synthetic pixel ${\bf x}[n]$ \cite{chan2009convex,ambikapathi2011chance,lin2014identifiability} has been defined as $\rho_n \triangleq\|{\bf s}[n]\| \in [1/\sqrt{N},1]$ (due to {\sf (A1)} and {\sf (A2)}); a larger index $\rho_n$ means higher pixel purity of  $\vcx[n]=\sum_{i=1}^Ns_{i}[n]\vca_i$. 
Each synthetic data set in the simulation is generated with a given purity level denoted as $\rho$, following the same data generation procedure as in \cite{chan2009convex,ambikapathi2011chance,lin2014identifiability},
where $\rho$ is a measure of mixing level of a data set. 
Specifically, 
a pool of sufficiently large number of synthetic data is first generated, and then from the pool, $L$ pixels with the {purity index} $\rho_n$ not greater than $\rho$ are randomly picked to form the desired data set with a purity level of $\rho$. 

In the above data generation, six endmembers (i.e., Jarsoite, Pyrope, Dumortierite, Buddingtonite, Muscovite, and Goethite)
with $224$ spectral bands randomly selected from the USGS library \cite{USGS2007} are used to generate $10000$ synthetic hyperspectral data ${\bf x}[n]$ (i.e., $N=6$, $M=224$, $L=10000$) with $\rho\in\{0.8,0.9,1\}$ and ${\rm SNR}\in \{20,25,30,35,40\}$ (dB).
The simulation results for $\phi_{en}$, $\phi_{ab}$, and computational time $T$
are displayed in Table \ref{tab:pur_snr_end}, where bold-face numbers correspond to the best performance (i.e., the smallest $\phi_{en}$, $\phi_{ab}$, and $T$) of all the HU algorithms under test for a specific $(\rho,{\rm SNR})$.

\begin{table*}[t]
 \scriptsize 
\caption{Performance comparison, in terms of $\phi_{en}$ (degrees), $\phi_{ab}$ (degrees), and average running time $T$ (seconds), of various HU algorithms for different data purity levels $\rho$ and SNRs{, where abundances are i.i.d. and Dirichlet distributed.}}\vspace{-0.3cm}

\begin{center}\label{tab:pur_snr_end}
\begin{tabular}{c|c|c|c|c|c|c|c|c|c|c|c|c}
  \hline\hline
  \multirow{3}{*}{Methods} & \multirow{3}{*}{~~$\rho$~~}& \multicolumn{5}{|c|}{$\phi_{en}$ (degrees)} & \multicolumn{5}{|c|}{$\phi_{ab}$ (degrees)} &  \multirow{3}{*}{$T$ (seconds)} 
  
  \\\cline{3-12}&  & \multicolumn{5}{|c|}{SNR (dB)} & \multicolumn{5}{|c|}{SNR (dB)} 
  \\\cline{3-12}&  & 20 & 25 & 30 & 35 & 40 & 20 & 25 & 30 & 35 & 40\\
\hline

\multirow{3}{*}{MVC-NMF}  
  & 0.8  & 	2.87   & 	2.31   & 	1.63   & 	1.23   & 	1.14 	  & 	13.18  &  	9.83   & 	7.14  &  	5.58   & 	5.04 \\
  & 0.9	  & 2.98   & 	1.78   & 	0.98   & 	0.57   & 	0.40 	  & 	12.67  &  	8.00   & 	4.64  &  	2.85  &  	2.16   & 1.68E+{2} \\
  & 1  & 	3.25   & 	1.91   & 	1.00   & 	0.52   & 	{\bf  0.21} 	  & 	12.30   & 	7.45   & 	4.14   & 	2.26   & 	{\bf 1.11}
 \\\hline
 
\multirow{3}{*}{MVSA}  
  & 0.8  & 	11.08   & 	6.23   & 	3.41   & 	1.87   & 	1.03   & 		21.78   & 	14.49   & 	8.71   & 	5.00   & 	2.85 \\
  & 0.9	  & 11.55   & 	6.46   & 	3.48   & 	1.90   & 	1.05 	  & 	21.89   & 	14.51   & 	8.63   & 	4.91  &  	2.82  &  3.54E+{0} \\
  & 1  & 	11.64  &  	6.51   & 	3.54   & 	1.93   & 	1.06   & 		21.67   & 	14.21   & 	8.49   & 	4.81   & 	2.72 
 	\\\hline
 
\multirow{3}{*}{MVES}  
  & 0.8	  & 10.66   & 	6.06   & 	3.39   & 	1.91   & 	1.16 	  & 	21.04   & 	14.21  &  	9.04   & 	5.51   & 	3.33 \\
  & 0.9	  & 10.17   & 	6.06   & 	3.48   & 	1.97   & 	1.12 	  & 	21.51   & 	14.48   & 	9.28   & 	5.69   & 	3.45  &  2.80E+{1} \\
  & 1  & 	9.95   & 	5.96 	  & 3.55 	  & 2.19   & 	1.30 	  & 	22.50   & 	15.34   & 	10.32   & 	7.11   & 	4.49 
 \\\hline
 
\multirow{3}{*}{SISAL}
  & 0.8	  & 3.97   & 	2.59   & 	1.59  & 	0.94   & 	0.53   & 		13.70   & 	8.68 	  & 5.22   & 	3.09   & 	1.80 \\
  & 0.9	  & 4.18   & 	2.70   & 	1.64   & 	0.95  &  	0.54 	  & 	13.55   & 	8.54   & 	5.11   & 	3.00  &  	1.75   & 2.59E+{0} \\
  & 1	  & 4.49   & 	2.87  &  	1.73   & 	0.99   & 	0.54 	  & 	13.40   & 	8.43   & 	5.03   & 	2.93   & 	1.66 
 	  \\\hline

\multirow{3}{*}{ ipMVSA}
     & { 0.8}	  & { 12.03}  & 	{ 7.05}   & 	{ 4.04}  & 	{ 2.02}  & 	{ 1.16}   & 	{	21.81}   & {	14.89} 	  & { 9.58}   & 	{ 5.32}   & { 2.23} \\
     & { 0.9}	  & { 12.63}   & { 	7.55}   & {4.04}  & {	2.05}  &  {	1.25} 	  & {	22.33}  & {	15.36}  & {	9.37}  & {	5.21}  &  { 3.31}   & { 9.86E{-1}} \\
     & { 1}	  & {  12.89}   & 	{ 7.80}  &  	{ 4.00}   & 	{ 2.13}   & 	{ 1.28} 	  & 	{ 22.16}   & 	{ 15.20}   & 	{ 9.06}   & 	{ 5.25}   & 	{ 3.28}
    	  \\\hline

\multirow{3}{*}{HyperCSI}     
   &  0.8  & 	{\bf 1.65}   & 	{\bf 1.20}  &  	{\bf 0.79}   & {\bf	0.54}   & 	{\bf 0.37}	  & {\bf	11.17}  & {\bf	7.35}  & {\bf	4.32}  & {\bf	2.65} & 	{\bf 1.64} \\
   &  0.9  & 	{\bf 1.37}   & 	{\bf 1.03}   & {\bf 0.64}  & {\bf	0.45}   & {\bf 0.32} 	  & 	{\bf 10.08}   & 	{\bf 6.40}   & {\bf	3.62}  & {\bf	2.25} & 	{\bf 1.38}   & {\bf \bf 5.39E{-2}} \\
   & 1  & 	{\bf 1.21}   & 	{\bf 0.83} 	  & {\bf 0.57}   & 	{\bf 0.39}   & 	0.27 	  & 	{\bf 9.28} 	  & {\bf 5.46}   & 	{\bf  3.23}   & 	{\bf 1.92}   & 	1.15 
   \\\hline

  \hline\hline
\end{tabular}
\end{center}
\vspace{-0.3cm}
\end{table*}

Some general observations from Table \ref{tab:pur_snr_end} are as follows. 
For fixed purity level $\rho$, all the algorithms under test perform better for larger SNR.
As expected, the proposed HyperCSI algorithm rightly performs better for higher data purity level $\rho$, but this performance behavior does not apply to the other {five} algorithms, perhaps because the non-convexity of the complicated simplex volume makes their performance behaviors more intractable w.r.t. different data purities.

Among the {five} existing benchmark Craig-criterion-based HU algorithms, MVC-NMF yields more accurate endmember estimates than the other algorithms
{at the highest computational cost, while ipMVSA is the most computationally efficient one with lower performance as a trade-off.}
Nevertheless, the proposed HyperCSI algorithm outperforms all the other {five} algorithms when the data are heavily mixed (i.e., $\rho = 0.8$) or moderately mixed (i.e., $\rho = 0.9$).
As for high data purity $\rho = 1$, the HyperCSI algorithm also performs best except for the case of $(\rho,{\rm SNR})=(1,40~{\rm dB})$.
On the other hand,
the computational efficiency of the proposed HyperCSI algorithm is about $1$ to $4$ orders of magnitude faster than the other {five} HU algorithms under test.
Note that the computational efficiency of the HyperCSI algorithm can be further improved by an order of $\setO(N)$ if parallel processing can be implemented in Step 4 (hyperplane estimation) and Step 7 (abundance estimation) in Table \ref{table_HyperCSI}.
{Moreover, ipMVSA is around 4 times faster than MVSA, but performs slightly worse than MVSA, perhaps because ipMVSA \cite{li2015minimumFastMVSA} does not adopt the hinge-type soft constraint (for noise resistance) as used in MVSA \cite{Li2008}.}

\vspace{-0.2cm}


\subsection{Performance Evaluation of HyperCSI Algorithm with Non-i.i.d., Non-Dirichlet and Sparse Abundances}\label{subsec:all_algo_2new_maps}

\begin{figure}[h]\vspace{.05cm}
\psfrag{a}[Bc][Bc]{(a)~Ground truth abundance maps of SYN1}
\psfrag{b}[Bc][Bc]{(b)~Ground truth abundance maps of SYN2}
\begin{center}

\includegraphics[width=.48\textwidth]{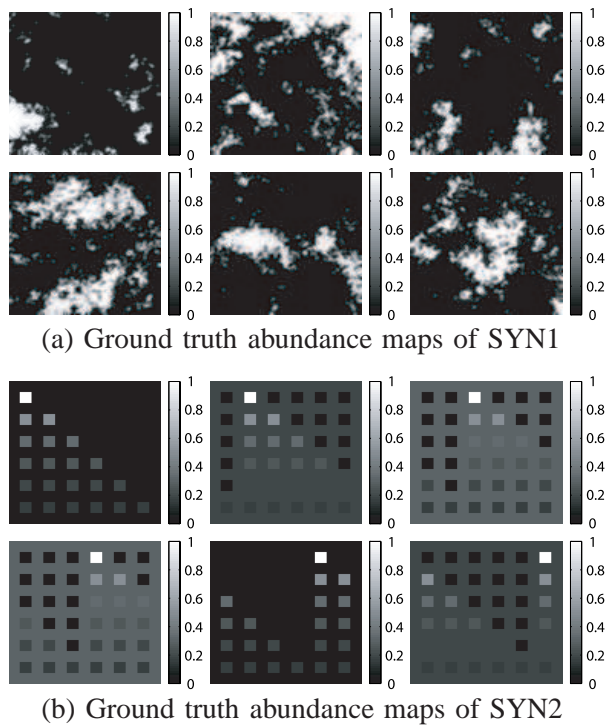}
\end{center}
\vspace{-0.95cm}
\caption{Two sets of sparse and spatially correlated abundance maps, where each subblock in subfigure (b) contains $10\times 10$ pixels.
    } \label{fig:two_new_maps}
\vspace{-0.35cm}
\end{figure}

{In practice, the abundance vectors ${\bf s}[n]$ may not be i.i.d. and seldom follow the Dirichlet distribution, and, moreover, the abundance maps often show large sparseness \cite{iordache2012totalvariation}.
In view of this, as considered in \cite{iordache2012totalvariation,chen2014nonlinear}, two sets of sparse and spatially correlated abundance maps displayed in Figure \ref{fig:two_new_maps} were used to generate two synthetic hyperspectral images, denoted as SYN1 ($L=100\times 100$) and SYN2 ($L=130\times 130$).
Then all the algorithms listed in Table \ref{parameters_table} are tested again with these two synthetic data sets for which the abundance vectors are obviously neither i.i.d. nor Dirichlet distributed.
}


{The simulation results,
in terms of $\phi_{en}$, $\phi_{ab}$, and computational time $T$, are shown in Table \ref{tab:pur_snr_end_2newmaps}, where bold-face numbers correspond to the best performance among the algorithms under test for a particular data set and a specific ${\rm SNR}\in \{20,25,30,35,40\}$ (dB).
As expected, for both data sets, all the algorithms perform better for larger SNR.
}

{One can see from Table \ref{tab:pur_snr_end_2newmaps} that for both data sets, HyperCSI yields more accurate endmember estimates than the other algorithms, except for the case of SNR$=40$ (dB).
As for abundance estimation,
HyperCSI performs best for SYN1, while MVC-NMF performs best for SYN2.
Moreover, among the five existing benchmark Craig-criterion-based
HU algorithms, ipMVSA and SISAL are the most computationally efficient ones.
However, in both data sets, the computational efficiency of the proposed HyperCSI algorithm is at least more than one order of magnitude faster than the other five algorithms.
These simulation results have demonstrated
the superior efficacy of the proposed HyperCSI algorithm over the other algorithms under test in both estimation accuracy and computational efficiency.
}

\begin{table*}[t]
\scriptsize 
\caption{Performance comparison, in terms of $\phi_{en}$ (degrees), $\phi_{ab}$ (degrees) and running time $T$ (seconds), of various HU algorithms using synthetic data SYN1 and SYN2 for different SNRs, where abundances are non-i.i.d., non-Dirichlet and sparse (see Figure \ref{fig:two_new_maps}).}\vspace{0.0cm}
\begin{center}\label{tab:pur_snr_end_2newmaps}
\renewcommand{\arraystretch}{1.1}
\begin{tabular}{c|c|c|c|c|c|c|c|c|c|c|c|c}

  \hline\hline
  \multirow{3}{*}{} & \multirow{3}{*}{Methods}& \multicolumn{5}{|c|}{$\phi_{en}$ (degrees)} & \multicolumn{5}{|c|}{$\phi_{ab}$ (degrees)} &  \multirow{3}{*}{$T$ (seconds)} 
  
  \\\cline{3-12}&  & \multicolumn{5}{|c|}{SNR (dB)} & \multicolumn{5}{|c|}{SNR (dB)} 
  \\\cline{3-12}&  & 20 & 25 & 30 & 35 & 40 & 20 & 25 & 30 & 35 & 40\\
\hline

\multirow{6}{*}{ SYN1} 
  & {MVC-NMF}	  & { 3.23}  & {	1.97}  & { 1.05} & { 0.55}  & {{\bf 0.25}}  & { 13.87}   & {8.51} & { 4.79}   & {	2.65}  & {	{\bf 1.34}} &{ 1.74E+{2}} \\\cline{2-13}
  
  & { MVSA}  & { 10.65}   & { 6.12}  & { 3.38} & { 1.88}  &  	{ 1.05}	  & { 22.93}   & {15.13}   & 	{ 9.34}   & {	5.52}  &  { 3.19}   & { 3.53E+{0}} \\\cline{2-13}
  
  & { MVES}	  & { 9.55}   & {	5.49}  &  { 3.60}   & { 1.96}   & 	{ 1.22}  & { 23.89}  & {17.35}   & { 14.49}   & 	{ 7.78}   & 	{ 5.66} & { 3.42E+{1}} \\\cline{2-13}
  
 & { SISAL} & { 4.43}  & { 2.89}   & { 1.81}  & {	1.18}  & 	{ 0.86}   & { 15.85}  & 	{ 10.39}	  & { 6.89}   & {	5.29}   & 	{4.65} &{ 2.66E+{0}}\\\cline{2-13}
 
 & { ipMVSA} & { 11.62}   & { 6.82}   & 	{ 3.38}  & { 2.01}   & { 1.05}   & { 24.05}   & { 16.28} 	  & { 9.34}   & {	5.98}   & 	{ 3.19} &{ 1.65E+{0}}\\\cline{2-13}

 & { HyperCSI}	  & {{\bf 1.55}}  & {{\bf	1.22}}  & {{\bf 0.79}} &{ {\bf 0.52}}  & { 0.35}   & {{\bf	12.03}}  & {{\bf 6.92}}& {{\bf 4.16}} & {{\bf 2.49}}   & 	{ 1.46} & { {\bf 5.56E{-2}}} \\	  
 	  \hline	  \hline
 	  
\multirow{6}{*}{ SYN2} 
  &{ MVC-NMF}	  & { 2.86}   & {	1.71}   & 	{ 0.97}  & { 0.54}   & 	{{\bf 0.23}}   & { 22.86}   & 	{{\bf 15.52}} & {{\bf 9.39}}   & 	{{\bf 5.27}}  & 	{{\bf 2.67}} &{ 2.48E+{2}} \\\cline{2-13}
  
  & { MVSA}  & { 10.21}   & 	{ 5.55}   & {	3.08}   & 	{ 1.71}  &  	{ 0.95}  & { 29.86}   & {22.72}   & { 15.57}   & {	9.78}  &  { 5.83}   & { 5.65E+{0}} \\\cline{2-13}
  
 & { MVES}	  & { 10.12}   & 	{ 5.19}  &  	{ 3.15}   & { 2.04}   & { 3.77}	  & { 29.43}   & { 22.13}   & {	15.66}   & 	{ 10.42}   & 	{13.17} & { 2.22E+{1}} \\\cline{2-13}
  
 & { SISAL}& { 3.25}   & { 2.18}   & { 1.48}  & {0.96}  & 	{ 0.63}   & {	24.79}   & 	{ 17.49}	  & { 11.51}   & 	{ 7.00}   & { 4.21} &{ 4.45E+{0}}\\\cline{2-13}
 
 & { ipMVSA}	  & { 11.34}   & 	{ 8.26}   & 	{ 3.34}  & 	{ 1.94}   & { 1.01}   & { 30.23}   & { 30.38} 	  & { 16.29}   & { 10.30}  & { 6.39} &{ 8.14E{-1}}\\\cline{2-13}

 & { HyperCSI}	  & {{\bf 1.48}}  & {{\bf 1.08}}  & { {\bf  0.71}} & { {\bf 0.44}}  & { 0.31}   & {{\bf 22.64}}  & { 15.98} &{ 11.10} & { 7.25}   & { 4.40} &  {{ \bf 7.48E{-2}}} \\	  
 	  \hline

  \hline\hline

\end{tabular}
\end{center}
\end{table*}

\section{Experiments {with AVIRIS Data}}
\label{sec:RealData}

\begin{figure}[htp!]
\begin{center}\ifconfver
\resizebox{0.4\linewidth}{!}{\includegraphics{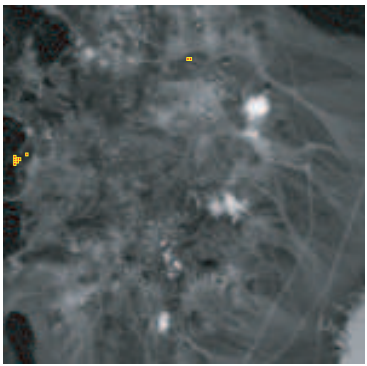}}
\vspace{-0.3cm}\else
    \resizebox{0.3\linewidth}{!}{\includegraphics{50th_band_outliers.eps}}
\fi
\end{center}\ifconfver
\else
    \vspace{-0.8cm}
\fi \caption{The subimage of the AVIRIS hyperspectral imaging data cube for
the 50th band, where the locations of the ten outliers identified by the RASF algorithm are marked with yellow color.} \label{Fig:real_image}
\end{figure}

In this section, the proposed HyperCSI algorithm along with two benchmark HU algorithms, i.e., the MVC-NMF algorithm \cite{miao2007endmember} developed based on Craig's criterion, and the VCA algorithm \cite{Nascimento2005} (in conjunction with the FCLS algorithm \cite{Heinz2001} for the abundance estimation) developed based on the pure-pixel assumption, are used to process the hyperspectral imaging data collected by the Airborne Visible/Infrared Imaging Spectrometer (AVIRIS) \cite{AVIRISrealdata} taken over the Cuprite mining site, Nevada, in 1997. 
We consider this mining site, not only because it has been extensively used for remote sensing experiments \cite{Clark2003}, but also because the available
{classification ground truth in \cite{Swayze1992,Swayze1997} (though which may have coregistration issue as it was obtained earlier than 1997, this ground truth has been widely accepted in the HU context)}
allows us to easily verify the experimental results.
The AVIRIS sensor is an imaging spectrometer with 224 channels (or spectral bands) that cover wavelengths ranging from $0.4$ to $2.5$ $\mu$m with an approximately 10-nm spectral resolution.
The bands with low SNR as well as those corrupted by water-vapor absorption (including bands 1-4, 107-114, 152-170, and 215-224) are removed from the original 224-band imaging data cube, and hence a total of $M=183$ bands is considered in our experiments.
Furthermore, the selected subscene of interest includes 150 vertical lines with 150 pixels per line, and its 50th band is shown in Figure \ref{Fig:real_image}(a), where the 10 pixels marked with yellow color are removed from the data set as they are outlier pixels identified by the robust affine set fitting (RASF) algorithm \cite{chan2013robust}.

The number $N$ of the minerals (i.e., endmembers) present in the selected subscene is estimated using a virtual dimensionality (VD) approach \cite{Chang2004}, i.e., the noise-whitened Harsanyi-Farrand-Chang (NWHFC) eigenvalue-thresholding-based algorithm with false-alarm probability $P_{FA} = 10^{-3}$. The obtained estimate is $\hat N=9$ and used in the ensuing experiments for all the three HU algorithms under test.      
%

\begin{figure}[h!]
\begin{center}
    \ifconfver\resizebox{1\linewidth}{!}{\includegraphics{Hyperplane_abundance_maps.eps}}
\else \resizebox{0.6\linewidth}{!}{\includegraphics{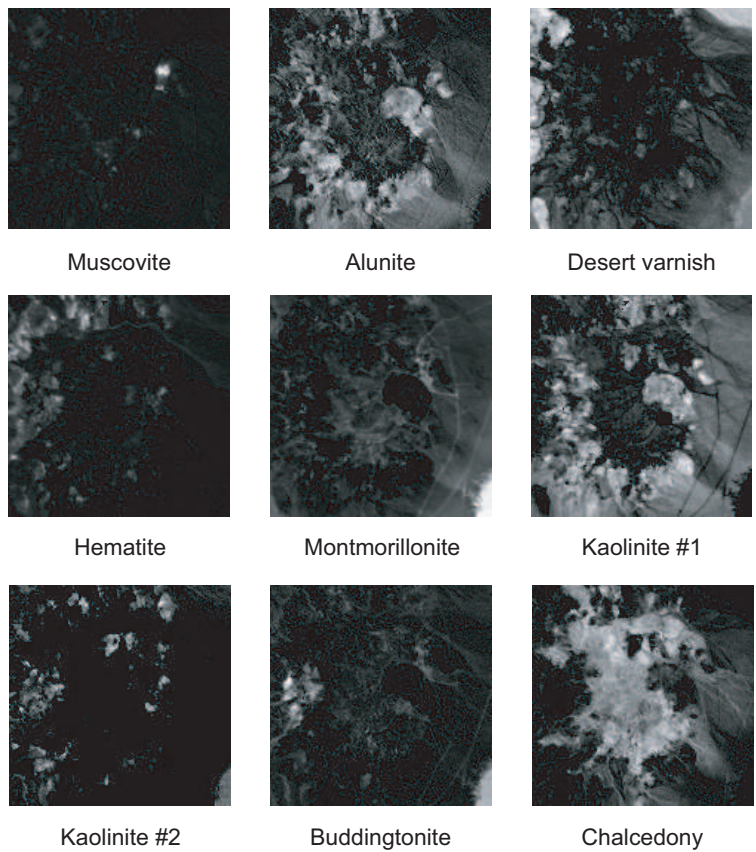}}
\fi
\end{center}
\vspace{-0.15cm}
\caption{The abundance maps of minerals estimated by HyperCSI algorithm.}\label{Fig:real_image_Hyp}
\end{figure}

The estimated abundance maps are visually compared with those reported in \cite{miao2007endmember,Nascimento2005,chan2009convex} as well as the ground truth reported in \cite{Swayze1992,Swayze1997}, so as to determine what minerals they are associated with.
The nine abundance maps obtained by the proposed HyperCSI algorithm are shown in Figure \ref{Fig:real_image_Hyp}, and they are identified as mineral maps of Muscovite, Alunite, Desert Varnish, Hematite, Montmorillonite, Kaolinite $\#$1, Kaolinite $\#$2, Buddingtonite, Chalcedony, respectively, as listed in Table \ref{tab:num_end}.
The minerals identified by MVC-NMF and VCA are also listed in Table \ref{tab:num_end}, where MVC-NMF also identifies nine distinct minerals, while only eight distinct minerals are retrieved by VCA, perhaps due to lack of pure pixels in the selected subscene or randomness involved in VCA. Owing to space limitation, their mineral maps are not shown here.
%
%

\begin{figure}[htp!]
\ifconfver
\begin{center}
\hspace{-0.78cm}
    \resizebox{1.04\linewidth}{!}{\includegraphics{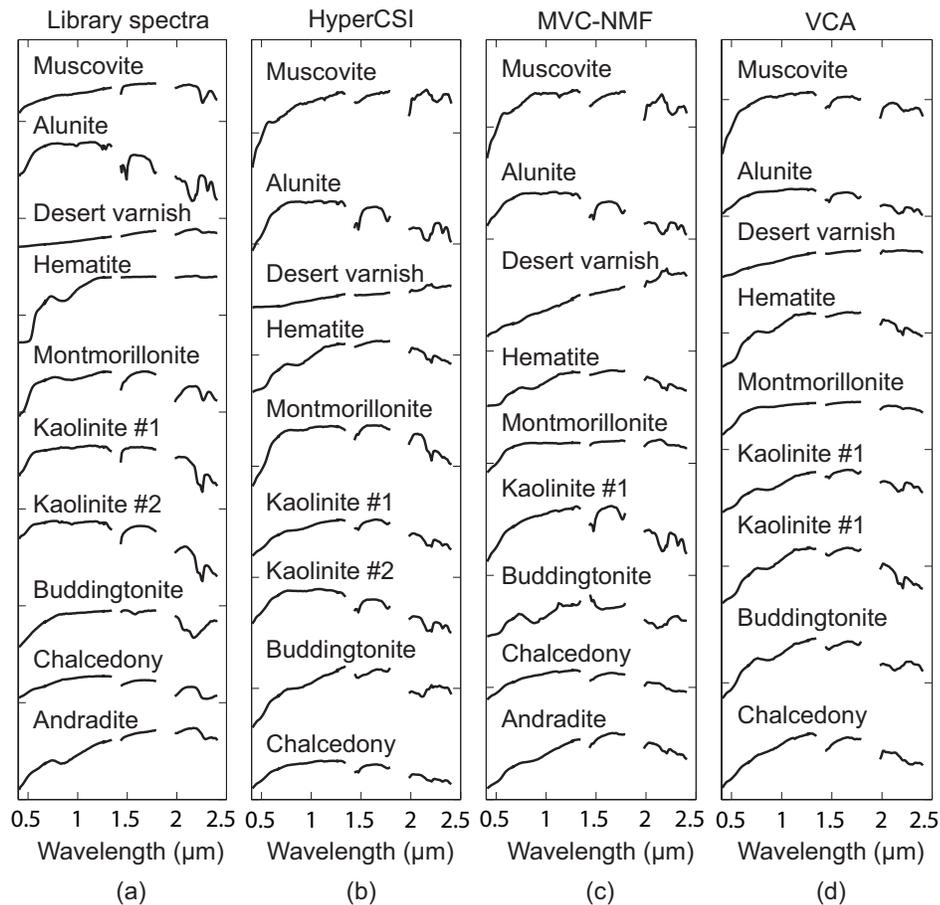}}\vspace{-0.3cm}
\end{center}
\else
\begin{center}
    \resizebox{0.8\linewidth}{!}{\includegraphics{ALL_signatures.eps}}
\end{center}
    \vspace{-0.5cm}
\fi \caption{(a) The endmember signatures taken from the USGS
library, and signatures of the endmember estimates obtained by (b) HyperCSI, (c) MVC-NMF and (d) VCA.} \label{Fig:real_endmember}
\end{figure}

The mineral spectra extracted by the three algorithms under test, along with their counterparts in the USGS library \cite{USGS2007}, are shown in Figure \ref{Fig:real_endmember}, where one can observe that the spectra extracted by the proposed HyperCSI algorithm hold a better resemblance to the library spectra.
For instance, the spectrum of Alunite extracted by HyperCSI shows much clearer absorption feature than MVC-NMF and VCA, in the bands approximately from $2.3$ to $2.5$ $\mu$m.
To quantitatively compare the endmember estimation accuracy among the three algorithms under test, the spectral angle distance between each endmember estimate $\hat{{\bf a}}$ and its corresponding library spectrum ${\bf a}$ serves as the performance measure and is defined as
\begin{equation}\label{SSE_a_realdata}
\phi = \arccos\left(\frac{{\bf a}^T\hat{{\bf a}}}{\|{\bf a}\| \cdot \|\hat{{\bf a}}\|}\right).\nonumber
\end{equation}
The values of ${\phi}$ associated with the endmember estimates for all the three algorithms under test are also shown in Table \ref{tab:num_end}, where the number in the parentheses is the value of ${\phi}$ associated with Kaolinite $\#$1 repeatedly classified by VCA.
One can see from Table \ref{tab:num_end} that the average of $\phi$ of the proposed HyperCSI algorithm is the smallest.
{The good performance of HyperCSI in endmember estimation intimates to that
the potential requirement of sufficient number (i.e., $N(N-1)=72$, in this experiment) of pixels lying close to the hyperplanes associated with the actual endmembers' simplex, has been met.
However, we are not too surprised with this observation, since the number of minerals present in one pixel ${\bf x}[n]$ is often small (typically, within five \cite{Ken14SPM_HU}), i.e., the abundance vector ${\bf s}[n]$ often shows sparseness \cite{iordache2012totalvariation} (cf. Figure \ref{Fig:real_image_Hyp}), indicating that a non-trivial portion of pixels are more likely to lie close to the boundary of the endmembers' simplex (note that $s_i[n]=0$ if, and only if, ${\bf x}[n]\in\setH_i$).}
Moreover, as the pure pixels may not be present in the selected subscene, as expected the two Craig-criterion-based HU algorithms (i.e., HyperCSI and MVC-NMF) outperform VCA in terms of endmember estimation accuracy.
On the other hand, in terms of the computation time $T$ as given in Table \ref{tab:num_end}, in spite of parallel processing not applied, the HyperCSI algorithm is 
{around 2.5 times faster than VCA (note that VCA itself only costs 0.31 seconds (out of the 5.40 seconds), and the remaining computation time is the cost of the FCLS)}
and almost four orders of magnitude faster than MVC-NMF.

\begin{table}[t]
\caption{The computational times $T$ (seconds) and spectral angle distance ${\phi}$ (degrees) between library spectra and endmembers estimated by HyperCSI, MVC-NMF, and VCA. The bold face numbers correspond to the smallest values of $\phi$ or $T$ among the three algorithms under test.}
\begin{center}

\begin{tabular}{c|c|c|c} \hline\hline
  & HyperCSI  & MVC-NMF & VCA \\\hline
Muscovite  & {\bf 3.03}  & 3.96 & 4.54 \\
Alunite & 7.48 & {\bf 6.23} & 6.57\\
Desert Varnish & 9.49 & {\bf 4.91} & 7.92\\
Hematite & 7.83 & 12.94 & {\bf 7.24}\\
Montmorillonite & {\bf 4.84} & 7.44 & 6.59\\
Kaolinite $\#$1 & 8.63  & {\bf 7.56} & 13.80 (11.71)\\
Kaolinite $\#$2 & {\bf 7.39} & -  & -\\
Buddingtonite & 6.55 & 8.16 & {\bf 6.46}\\
Chalcedony & {\bf 5.92} & 7.97 & 8.25\\
Andradite & - & {\bf 7.43}  & - \\\hline
Average $\phi$ (degrees) & {\bf 6.80}  &	7.40  &	8.12 \\\hline
$T$ (seconds) & {\bf 0.12} & 988.67 
 & 5.40 \\
\hline\hline
\end{tabular}\label{tab:num_end}
\end{center}
\vspace{-0.6cm}
\end{table}


\section{Conclusions}
\label{sec:Conclusions}

Based on the hyperplane representation for a simplest simplex,
we have presented an effective and computationally efficient Craig-criterion-based HU algorithm, called HyperCSI algorithm, given in Table \ref{table_HyperCSI}. The proposed HyperCSI algorithm has the following remarkable characteristics:
\begin{itemize}
\item It never requires the presence of pure pixels in the data.

\item It is reproducible without involving random initialization.

\item It only involves simple linear algebraic computations, and suitable for parallel implementation.
{Its computational complexity (without using parallel implementation) is $\mathcal{O}(N^2L)$, which is also the complexity of some state-of-the-art pure-pixel-based HU algorithms.}

\item It estimates Craig's minimum-volume simplex by finding only
$N(N-1)$ pixels (regardless of the data length $L$) from the data set for the construction of the associated hyperplanes, without involving any simplex volume computations, thereby accounting for its high computational efficiency in endmember estimation.

\item The estimated endmembers are guaranteed non-negative, and the
identified simplex was proven to be both Craig's simplex and true
endmembers' simplex w.p.1. as $L\rightarrow\infty $ for the noiseless case. 

\item The abundance estimation is readily fulfilled by a closed-form expression, and thus is computationally efficient.  
\end{itemize}

Some simulation results were presented to demonstrate the analytic results on the asymptotic endmember identifiability of the proposed HyperCSI algorithm, and its superior efficacy over some state-of-the-art Craig-criterion-based HU algorithms in both solution accuracy and computational efficiency. Finally, the proposed HyperCSI algorithm was tested using {AVIRIS} hyperspectral data to show its applicability.

    \section*{Appendix}
    \renewcommand{\thesubsection}{\Alph{subsection}}

\subsection{Proof of Theorem~\ref{prop:AI}}
\label{proofsec:AI}

For a fixed $i\in\setI_N$, one can see from \eqref{eq:region_def} that 
$\setR_{k}^{(i)} \cap \setR_{\ell}^{(i)} = \emptyset ,~\forall~k\neq \ell$, implying that the $N-1$ pixels $\vcp_k^{(i)}$, $\forall~k\in\setI_{N-1}$, identified by solving \eqref{eq:multiple_vector} must be distinct.
Hence, it suffices to show that $\setP$ is affinely independent w.p.1 for any $\setP \triangleq \{\vcp_1,\ldots,\vcp_{N-1} \} \subseteq \setX$ that satisfies
\begin{equation}\vspace{-0.15cm}
\vcp_{k} \neq \vcp_{\ell},~\textrm{for all}~1\leq k<\ell\leq N-1.\label{eq:distinct}
\end{equation}
Then, as $ \vcp_{k} \in \setX,~\forall~k\in\setI_{N-1} $, we have from {\sf (A4)} and \eqref{eq:distinct} that there exist i.i.d Dirichlet distributed random vectors $\{\vcs_1,\ldots,\vcs_{N-1}\} \subseteq \mathsf{dom}~f$ such that (cf. \eqref{eq:pprrox})
\begin{equation}\vspace{-0.2cm}
\vcp_k = [\boldsymbol{\alpha}_1\cdots\boldsymbol{\alpha}_N]~ \vcs_k,~~~~\textrm{for all}~k\in\setI_{N-1}.\label{eq:prop3_222}
\end{equation}
For ease of the ensuing presentation, let $\Pr\{\cdot\}$ denote the probability function and define the following events:
\ifconfver
     \begin{Ventry}{~~$\textsf{E3}^{(k)}$~~}
     \item [~~$\textsf{E1}$] The set $\setP$ is affinely dependent.
     
     \item [~~$\textsf{E2}$] The set $\{\vcs_1,\ldots,\vcs_{N-1}\}$ is affinely dependent.
     
     \item [~~$\textsf{E3}^{(k)}$] $\vcs_k \in {\sf aff}\left\{ \{\vcs_1,\ldots,\vcs_{N-1}\} \setminus \{\vcs_k\} \right\}$, $\forall~k\in\setI_{N-1}$.
     \end{Ventry}
\else
    \begin{Ventry}{~~~$\textsf{E3}^{(k)}$~~~~~~~~~~~~}
    \item [~~~~~~~~~~~~$\textsf{E1}$] The set $\setP$ is affinely dependent.
    
    \item [~~~~~~~~~~~~$\textsf{E2}$] The set $\{\vcs_1,\ldots,\vcs_{N-1}\}$ is affinely dependent.
    
    \item [~~~~~~~~~~~~$\textsf{E3}^{(k)}$] $\vcs_k \in {\sf aff}\left\{ \{\vcs_1,\ldots,\vcs_{N-1}\} \setminus \{\vcs_k\} \right\}$, $\forall~k\in\setI_{N-1}$.
    \end{Ventry}
\fi
Then, to prove that $\setP_i$ is affinely independent w.p.1, it suffices to prove $\Pr\{ \textsf{E1} \} = 0$.

Next, let us show that {\sf E1} implies {\sf E2}.
Assume {\sf E1} is true.
Then $\vcp_k \in {\sf aff} \{ \setP \setminus \{\vcp_k\} \}$ for some $k\in\setI_{N-1}$.
Without loss of generality, let us assume $k=1$.
Then,
\begin{equation}\vspace{-0.15cm}
\vcp_1 = \theta_2 \cdot \vcp_2 + \cdots + \theta_{N-1} \cdot \vcp_{N-1},\label{eq:prop3_333}
\end{equation}
for some $\theta_i,i=2,\dots,N-1$, satisfying
\begin{equation}\vspace{-0.1cm}
\theta_2 + \cdots + \theta_{N-1} = 1.\label{eq:prop3_444}
\end{equation}
By substituting \eqref{eq:prop3_222} into \eqref{eq:prop3_333}, we have
\begin{equation}
[\boldsymbol{\alpha}_1,\ldots,\boldsymbol{\alpha}_N]~ \vcs_1
=
[\boldsymbol{\alpha}_1,\ldots,\boldsymbol{\alpha}_N]~ \vct,\label{eq:prop3_555}
\end{equation}
where $\vct \triangleq  \sum_{m=2}^{N-1} ~ \theta_m \cdot\vcs_m $.
For notational simplicity, let $[\vcu]_{1:N-1} \triangleq [u_1,\ldots, u_{N-1}]^T$ for any given vector $\vcu=[ u_1,\ldots,u_N ]^T$.
Then, from the facts of ${\bf 1}^T_N\vct  =1$ (by \eqref{eq:prop3_444}) and ${\bf 1}^T_N \vcs_1 =1$, \eqref{eq:prop3_555} can be rewritten as
\begin{align}\label{eq:prop3_201503111144}
{\bf \Theta} ~ [\vcs_1]_{1:N-1} 
= {\bf \Theta} ~ [\vct]_{1:N-1},
\end{align}
where ${\bf \Theta} \triangleq [\boldsymbol{\alpha}_1-\boldsymbol{\alpha}_N ,\ldots, \boldsymbol{\alpha}_{N-1}-\boldsymbol{\alpha}_N]$.
As $\{ \boldsymbol{\alpha}_1,\ldots,\boldsymbol{\alpha}_N \}$ is affinely independent (by {\sf (A3)}), the matrix ${\bf \Theta}$ is of full column rank \cite{CVX2004}, implying that $[\vcs_1]_{1:N-1} = [\vct]_{1:N-1}$ (by \eqref{eq:prop3_201503111144}).
Then, by the facts of ${\bf 1}^T_N\vct  =1$ and ${\bf 1}^T_N \vcs_1 =1$, one can readily come up with $\vcs_1 = \vct = \sum_{m=2}^{N-1} ~ \theta_m \cdot\vcs_m$, or, equivalently, $\vcs_1 \in {\sf aff}\{ \vcs_2,\ldots,\vcs_{N-1} \}$ (by \eqref{eq:prop3_444}), implying that {\sf E2} is true \cite{CVX2004}.
Thus we have proved that {\sf E1} implies {\sf E2}, and hence
\begin{equation}\vspace{-0.1cm}
\Pr\{ \textsf{E1} \} \leq \Pr\{ \textsf{E2} \}.\label{eq:prop3_e1e2}
\end{equation}

As Dirichlet distribution is a continuous multivariate distribution \cite{johnson2002continuous} for a random vector ${\bf s}\in \mathbb{R}^N$ to satisfy {\sf (A1)}-{\sf (A2)} with an $(N-1)$-dimensional domain, any given affine hull $\setA\subseteq\mathbb{R}^N$ with affine dimension $P$ must satisfy \cite{frigyik2010introduction}
\begin{equation}\vspace{-0.1cm}
\Pr\{~ \vcs\in\setA ~\} = 0,~\textrm{if}~P<N-1.\label{eq:prop3_aaa}
\end{equation}
Moreover, as $\{\vcs_1,\ldots,\vcs_{N-1}\}$ are i.i.d. random vectors and the affine hull ${\sf aff}\left\{ \{\vcs_1,\ldots,\vcs_{N-1}\} \setminus \{\vcs_k\} \right\}$ must have affine dimension $P< N-1$, we have from \eqref{eq:prop3_aaa} that 
\begin{equation}\vspace{-0.15cm}
\Pr\{\textsf{E3}^{(k)}\} = 0,~\textrm{for all}~k\in\setI_{N-1}.\label{eq:prop3_bbb}
\end{equation}
Then we have the following inferences:
\ifconfver
     \begin{align*}
     0 &\leq \Pr\{ \textsf{E1} \} \leq \Pr\{ \textsf{E2} \}~~\textrm{(by \eqref{eq:prop3_e1e2})}\nonumber\\
     &=\Pr\{ \cup_{k=1}^{N-1}~ \textsf{E3}^{(k)} \}~~\textrm{(by the definitions of $\textsf{E2}$ and $\textsf{E3}^{(k)}$)}\nonumber\\
     &\leq \sum_{k=1}^{N-1}~\Pr\{ \textsf{E3}^{(k)} \}~=~0,~~\textrm{(by the union bound and \eqref{eq:prop3_bbb})}\nonumber
     \end{align*}
\else
    \begin{align*}
    0 &\leq \Pr\{ \textsf{E1} \} \leq \Pr\{ \textsf{E2} \}~~~~~~\textrm{(by \eqref{eq:prop3_e1e2})}\nonumber\\
    &=\Pr\{ \cup_{k=1}^{N-1}~ \textsf{E3}^{(k)} \}~~~~~~\textrm{(by the definitions of $\textsf{E2}$ and $\textsf{E3}^{(k)}$)}\nonumber\\
    &\leq \sum_{k=1}^{N-1}~\Pr\{ \textsf{E3}^{(k)} \}~=~0,~~~~~~\textrm{(by the union bound and \eqref{eq:prop3_bbb})}\nonumber
    \end{align*}
\fi
i.e., $\Pr\{ \textsf{E1} \} = 0$. Therefore, the proof is completed.\hfill$\blacksquare$

\subsection{Proof of Theorem~\ref{thm:identifiability}}
\label{proofsec:THM_identifiability}

It can be seen from \eqref{eq:dirich_pdf} that the p.d.f. of Dirichlet distribution satisfies
\begin{equation}\label{B09100455}
f(\vcs) = \frac{\Gamma(\gamma_0)}{\prod_{i=1}^N \Gamma(\gamma_i) } \cdot \prod_{i=1}^N s_i^{\gamma_i -1} >0,~\forall~\vcs\in \mathsf{dom}~f.
\end{equation}
Moreover, by the facts of ${\bf A}{\bf e}_i={\bf a}_i$ and 
\[
\mathsf{dom}~f = \{\vcs\in\mathbb{R}^N_{++}~\big| ~{\bf 1}^T_N\vcs = 1\} = \intr~ \conv\{ \vce_1,\ldots,\vce_N \},
\]  
where $\intr~\setU$ denotes the interior of a set $\setU$,
the linear mapping (i.e., $\vcx={\bf A}\vcs$) of the abundance domain $\mathsf{dom}~f$ full fills the interior of the true endmembers' simplex $\conv\{ {\bf a}_1,\ldots,{\bf a}_N \}$, namely
\begin{align}\label{B09100502}
\{ \vcx={\bf A}\vcs ~\big|~ \vcs\in \mathsf{dom}~f \}
= \intr ~ \conv\{ {\bf a}_1,\ldots,{\bf a}_N \}.
\end{align}
Then, from \eqref{B09100455}-\eqref{B09100502} and {\sf (A4)}, it can be inferred that
\[
\Pr\{ \conv\{\vcx[1],\ldots,\vcx[L]\}|_{L\rightarrow \infty} = \intr~\conv\{ {\bf a}_1,\ldots,{\bf a}_N \} \} =1,
\]
which, together with the fact that the affine mapping (cf. \eqref{eq:pprrox}) preserves the geometric structure of $\{\vcx[1],\ldots,\vcx[L]\}$ \cite{Chan2007}
(note that ${\bf C}^T{\bf C}={\bf I}_{N-1}$),
further implies
\begin{equation}\label{B09101940}
\Pr\{~ \conv\setX = \intr~\conv\{ \boldsymbol{\alpha}_1,\ldots,\boldsymbol{\alpha}_N \} ~\} =1,
\end{equation}
where $\setX = \{ \tilde{{\bf x}}[1],\ldots,\tilde{{\bf x}}[L] \}|_{L\rightarrow \infty}$ throughout the ensuing proof. 
It can be inferred from \eqref{B09101940} that there is always a pixel $\tilde{\vcx}[n] \in \setX$ that can be arbitrarily close to the extreme point $\boldsymbol{\alpha}_i$ of the simplex $\conv\{ \boldsymbol{\alpha}_1,\ldots,\boldsymbol{\alpha}_N \}$, i.e., for all $i\in\setI_N$, 
\begin{equation}\label{09102124}
\Pr\{~ \setB( \boldsymbol{\alpha}_i, \epsilon ) \cap \setX \neq \emptyset ~\} =1,~\textrm{for any}~\epsilon>0.
\end{equation}

Let $\MVES(\setU)$ denote the set of all minimum-volume enclosing simplexes of $\setU$ (i.e., Craig's simplex containing the set $\setU$).
{Then, one can infer from the convexity of a simplex that (cf. \cite[Equation (32)]{lin2014identifiability})}
\begin{equation}\label{B09102130}
\MVES (\setX) = \MVES(\conv \setX).
\end{equation}
Moreover, by the fact that any simplex $\setT$ must also be a closed set and the fact that the closure of $\intr(\conv\{ \boldsymbol{\alpha}_1,\ldots,\boldsymbol{\alpha}_N \})$ is exactly $\conv\{ \boldsymbol{\alpha}_1,\ldots,\boldsymbol{\alpha}_N \}$, it can be seen that
$\conv\{ \boldsymbol{\alpha}_1,\ldots,\boldsymbol{\alpha}_N \} \subseteq\setT$ if and only if $\intr(\conv\{ \boldsymbol{\alpha}_1,\ldots,\boldsymbol{\alpha}_N \}) \subseteq\setT$ \cite{apostol1974mathematical}, and hence
\begin{equation}\label{B09102132}
\MVES( \conv\{ \boldsymbol{\alpha}_1,\ldots,\boldsymbol{\alpha}_N \} ) = \MVES( \intr~\conv\{ \boldsymbol{\alpha}_1,\ldots,\boldsymbol{\alpha}_N \} ).
\end{equation}
Thus, it can be inferred from \eqref{B09101940}, \eqref{B09102130} and \eqref{B09102132} that
\begin{equation}\label{B09102200}
\Pr\{~ \MVES( \setX ) = \MVES( \conv\{ \boldsymbol{\alpha}_1,\ldots,\boldsymbol{\alpha}_N \} )~\} =1.
\end{equation}
As $\conv\{ \boldsymbol{\alpha}_1,\ldots,\boldsymbol{\alpha}_N \}$ itself is a simplex, $\MVES( \conv\{ \boldsymbol{\alpha}_1,\ldots,\boldsymbol{\alpha}_N \} ) = \{ \conv\{ \boldsymbol{\alpha}_1,\ldots,\boldsymbol{\alpha}_N \} \}$, which together with \eqref{B09102200} yields 
\begin{equation}\label{B09102217}
\Pr\{~ \MVES( \setX ) = \{ \conv\{ \boldsymbol{\alpha}_1,\ldots,\boldsymbol{\alpha}_N \} \} ~\} =1.
\end{equation}
In other words, we have proved that the Craig's minimum-volume simplex
is always the true endmembers' simplex $\conv\{ \boldsymbol{\alpha}_1,\ldots,\boldsymbol{\alpha}_N \}$.
To complete the proof of Theorem \ref{thm:identifiability}, it suffices to show that the true endmembers' simplex is always identical to the simplex identified by the HyperCSI algorithm, i.e., for all $i\in\setI_N$,
\begin{equation}\label{B09102233}
\Pr\{~   \hat{\boldsymbol{\alpha}}_i \in \setB( \boldsymbol{\alpha}_i, \epsilon ) ~\} =1,~\textrm{for any}~\epsilon>0,
\end{equation}
where $\hat{\boldsymbol{\alpha}}_1,\ldots,\hat{\boldsymbol{\alpha}}_N$ are the estimated DR endmembers using HyperCSI algorithm.

To this end, let us first show that, for all $i\in\setI_N$,
\begin{equation}\label{09111648}
\Pr\{~ \tilde{\boldsymbol{\alpha}}_i \in \setB( \boldsymbol{\alpha}_i, \epsilon ) ~\} =1,~\textrm{for any}~\epsilon>0,
\end{equation}
where $\{ \tilde{\boldsymbol{\alpha}}_1,\ldots,\tilde{\boldsymbol{\alpha}}_N \}$ are the purest pixels identified by SPA (cf. Step 2 in Table \ref{table_HyperCSI}). 
However, directly proving \eqref{09111648} is difficult due to the post-processing involved in SPA (cf. Algorithm 4 in \cite{SPAarora2012practical}).
In view of this, let $\tilde{{\bf x}}[\ell_1],\ldots,\tilde{{\bf x}}[\ell_N]$ be those pixels identified by SPA {\it before} post-processing. Because the post-processing is nothing but to obtain the purest pixel $\tilde{\boldsymbol{\alpha}}_i$ by iteratively pushing each $\tilde{{\bf x}}[\ell_i]$ away from the hyperplane $\aff\{ \tilde{{\bf x}}[\ell_j]~\big|~j\neq i \}$ \cite{SPAarora2012practical}, we have the following simplex volume inequalities
\begin{equation}\label{09111648vol}
V(\tilde{{\bf x}}[\ell_1],\ldots,\tilde{{\bf x}}[\ell_N]) \leq V( \tilde{\boldsymbol{\alpha}}_1,\ldots,\tilde{\boldsymbol{\alpha}}_N ) \leq V( \boldsymbol{\alpha}_1,\ldots,\boldsymbol{\alpha}_N ),
\end{equation}
where the last inequality is due to $\tilde{\boldsymbol{\alpha}}_i \in \setX \subseteq \conv\{ \boldsymbol{\alpha}_1,\ldots,\boldsymbol{\alpha}_N \}$.
Hence, by \eqref{09111648vol}, to prove \eqref{09111648}, it suffices to show that, for all $i\in\setI_N$,
\begin{equation}\label{09131125}
\Pr\{~ \tilde{{\bf x}}[\ell_i] \in \setB( \boldsymbol{\alpha}_i, \epsilon ) ~\} =1,~\textrm{for any}~\epsilon>0.
\end{equation}
However, 
the SPA before post-processing (cf. Algorithm 4 in \cite{SPAarora2012practical}) is exactly the same as the TRIP algorithm (cf. Algorithm 2 in \cite{ambikapathi2011two}), and it has been proven in \cite[Lemma 3]{ambikapathi2011two} that
\eqref{09102124} straightforwardly yields \eqref{09131125} for $\epsilon=0$;
note that the condition ``\eqref{09102124} with $\epsilon=0$" is equivalent to the pure-pixel assumption required in \cite[Lemma 3]{ambikapathi2011two}.
One can also show that \eqref{09102124} yields \eqref{09131125} for any $\epsilon>0$, and the proof basically follows the same induction procedure as in the proof of \cite[Lemma 3]{ambikapathi2011two} and is omitted here for conciseness.
Then, recalling that \eqref{09131125} is a sufficient condition for \eqref{09111648} to hold, we have proven \eqref{09111648}.

By the fact that ${\bm v}_i$ is a continuous function (cf. \eqref{eq:formula_of_bi}) and by \eqref{eq:bi_tilde_def} and \eqref{eq:region_def}, we see that
\begin{equation}\label{B09141538}
\begin{split}
       & \lim_{ \tilde{\boldsymbol{\alpha}}_i \rightarrow \boldsymbol{\alpha}_i,~\forall i} ~~\tilde{\vcb}_i
       ~~~ =~ {\bm v}_i(\boldsymbol{\alpha}_1, \ldots,\boldsymbol{\alpha}_N ) ~=~ {\vcb}_i,\\
       & \lim_{ \tilde{\boldsymbol{\alpha}}_i \rightarrow \boldsymbol{\alpha}_i,~\forall i} ~~\setR^{(i)}_k
       ~ =~ \setR^{(i)}_k(\boldsymbol{\alpha}_1, \ldots,\boldsymbol{\alpha}_N ).
\end{split}
\end{equation}
Moreover,
we have from \eqref{B09100455}, \eqref{09111648}, \eqref{B09141538} and $L\rightarrow\infty$ that the pixel $\vcp^{(i)}_k$ identified by \eqref{eq:multiple_vector} can be arbitrarily close to $\setH_i$.
Furthermore, by Theorem \ref{prop:AI}, we have that the vectors $\{\vcp_{1}^{(i)},\ldots,\vcp_{N-1}^{(i)}\}$ are
not only arbitrarily close to $\setH_i$,
but also affinely independent w.p.1,
which together with Proposition \ref{proposition:bi_formula_with_zero} implies that the estimated $\hat{\vcb}_i$ (cf. \eqref{eq:formula_of_bi_hat_hat}) is arbitrarily close to the true ${\vcb}_i$ w.p.1,
provided that the outward-pointing normal vectors $\hat{\vcb}_i$ and ${\vcb}_i$
have
the same norm without loss of generality.
Then, from \eqref{B09100455}, \eqref{eq:compute_inner_product_constant_hat}, and the premises of $L\rightarrow\infty$ and $c=1$,
it can be inferred that
the estimated hyperplane
$\hat{\setH}_i \equiv \setH_i(\hat{{\bf b}}_i,\hat{h}_i/c) = \setH_i(\hat{{\bf b}}_i,\hat{h}_i)$ is arbitrarily close to the true
$\setH_i \equiv \setH_i({\bf b}_i,h_i) $ (cf. \eqref{eqn:prop1_1}); precisely, we have
\begin{equation}\label{B09141558}
\Pr\{~   [\hat{\vcb}_i^T,\hat{h}_i]^T \in \setB( [\vcb_i^T,h_i]^T, \epsilon ) ~\} =1,~\textrm{for any}~\epsilon>0.
\end{equation}
Consequently, by comparing the formulas of $\boldsymbol{\alpha}_i$ (cf. \eqref{eqn:prop1_3}) and $\hat{\boldsymbol{\alpha}}_i$ (cf. \eqref{eqn:estimated_endmembers}), we have, from $c=1$ and \eqref{B09141558}, that $\hat{\boldsymbol{\alpha}}_i$ is always arbitrarily close to $\boldsymbol{\alpha}_i$, i.e., \eqref{B09102233} is true for all $i\in\setI_N$, and hence the proof of Theorem \ref{thm:identifiability} is completed.
\hfill$\blacksquare$

\bibliographystyle{IEEEtran}


\end{document}